\DocumentMetadata{}

\documentclass[sigconf,screen]{acmart}

\usepackage{multirow}
\usepackage{graphicx}
\usepackage{graphics}
\usepackage{subcaption}
\usepackage{dcolumn}
\usepackage{tabularray}

\AtBeginDocument{%
  \providecommand\BibTeX{{%
    \normalfont B\kern-0.5em{\scshape i\kern-0.25em b}\kern-0.8em\TeX}}}

\copyrightyear{2025}
\acmYear{2025}
\setcopyright{acmlicensed}\acmConference[CHI '25]{CHI Conference on Human Factors in Computing Systems}{April 26-May 1, 2025}{Yokohama, Japan}
\acmBooktitle{CHI Conference on Human Factors in Computing Systems (CHI '25), April 26-May 1, 2025, Yokohama, Japan}
\acmDOI{10.1145/3706598.3714189}
\acmISBN{979-8-4007-1394-1/25/04}

\begin{document}

\title[Multimodal Reflection Nudges for Deliberativeness]{Enhancing Deliberativeness: Evaluating the Impact of Multimodal Reflection Nudges}

\author{ShunYi Yeo}
\email{yeoshunyi.sutd@gmail.com}
\affiliation{%
  \institution{Singapore University of Technology and Design}
  \city{Singapore}
  \country{Singapore}
}

\author{Zhuoqun Jiang}
\email{zhuoqun\_jiang@mymail.sutd.edu.sg}
\affiliation{%
  \institution{Singapore University of Technology and Design}
  \city{Singapore}
  \country{Singapore}
}

\author{Anthony Tang}
\email{tonyt@smu.edu.sg}
\affiliation{%
  \institution{Singapore Management University}
  \city{Singapore}
  \country{Singapore}
}

\author{Simon Tangi Perrault}
\email{perrault.simon@gmail.com}
\affiliation{%
  \institution{Singapore University of Technology and Design}
  \city{Singapore}
  \country{Singapore}
}

\renewcommand{\shortauthors}{Yeo et al.}

\begin{abstract}
Nudging participants with text-based reflective nudges enhances deliberation quality on online deliberation platforms. The effectiveness of multimodal reflective nudges, however, remains largely unexplored. Given the multi-sensory nature of human perception, incorporating diverse modalities into self-reflection mechanisms has the potential to better support various reflective styles. This paper explores how presenting reflective nudges of different types (direct: persona and indirect: storytelling) in different modalities (text, image, video and audio) affects deliberation quality. We conducted two user studies with 20 and 200 participants respectively. The first study identifies the preferred modality for each type of reflective nudges, revealing that text is most preferred for persona and video is most preferred for storytelling. The second study assesses the impact of these modalities on deliberation quality. Our findings reveal distinct effects associated with each modality, providing valuable insights for developing more inclusive and effective online deliberation platforms.
\end{abstract}

\begin{CCSXML}
<ccs2012>
   <concept>
       <concept_id>10003120.10003121.10003129</concept_id>
       <concept_desc>Human-centered computing~Interactive systems and tools</concept_desc>
       <concept_significance>100</concept_significance>
       </concept>
   <concept>
       <concept_id>10003120.10003121.10011748</concept_id>
       <concept_desc>Human-centered computing~Empirical studies in HCI</concept_desc>
       <concept_significance>100</concept_significance>
       </concept>
 </ccs2012>
\end{CCSXML}

\ccsdesc[500]{Human-centered computing}
\ccsdesc[300]{Empirical studies in collaborative and social computing}

\keywords{deliberation, deliberativeness, deliberative quality, internal reflection, online deliberation, public discussions, nudges, indirect reflector, direct reflector, reflection, self-reflection, multimedia, multi-modality, large language model, civic engagement}

\begin{teaserfigure}
\centering
  \includegraphics[width=\textwidth]{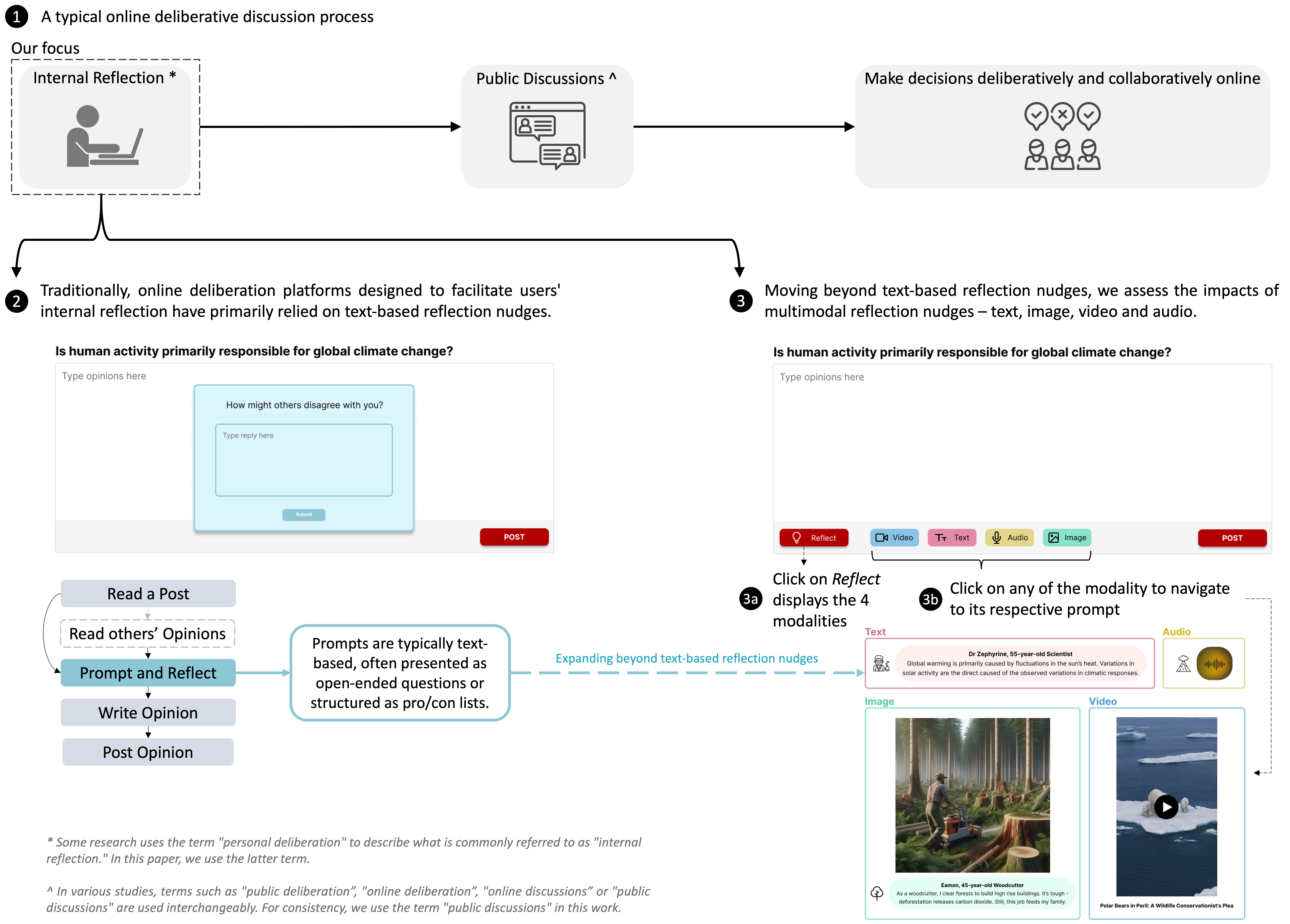}
  \caption{\textbf{1:} In online deliberation, the quality of collective discussions is directly influenced by the depth of individual reflection preceding them; \textbf{2:} To enhance users' reflection, text-based reflective nudges are commonly employed; \textbf{3:} To support various reflecting styles of individuals, we move beyond traditional text-based reflective nudges to multimodal reflective nudges; \textbf{3a and 3b:} We present \textit{Reflect}, a simple interface nudge powered by AI that showcases an array of distinct modalities, each guiding users' self-reflection through different presentations of Video (blue), Text (pink), Audio (yellow) and Image (green).}
  \label{fig: teaser}
  \Description{1: In online deliberation, the quality of collective discussions is directly influenced by the depth of individual reflection preceding them. 2: To enhance users' reflection, text-based reflective nudges are commonly employed. 3: To support various reflecting styles of individuals, we move beyond traditional text-based reflective nudges to multimodal reflective nudges. 3a and 3b: We present Reflect, a simple interface nudge powered by AI that showcases an array of distinct modalities, each guiding users' self-reflection through different presentations of Video (blue), Text (pink), Audio (yellow) and Image (green).}
\end{teaserfigure}



\maketitle

\section{Introduction}

In the realm of online deliberation, where public discussion is often perceived as inflammatory and hyperbolic~\cite{diakopoulos2011towards}, reflection is recognized as a critical element for enhancing the quality of discourse~\cite{muradova2021seeing, dryzek2002deliberative, goodin2000democratic, chambers2003deliberative, goodin2003does}. Online deliberation platforms are spaces for users to engage in thoughtful discussions on societal issues~\cite{jacobs2009talking}, often requiring participants to reflect on their own views~\cite{zhang2021nudge, goodin2003does, bohman2000public, goodin2000democratic} before collaboratively engaging with others to reach consensus~\cite{dekker2015contingency}. By fostering reflection, these platforms aim to improve the quality of individual contributions, leading to more meaningful and well-rounded deliberation.

Current approaches to reflection within online deliberation have predominantly relied on textual methods, such as creating pro/con lists~\cite{kriplean2012supporting}, engaging in perspective-taking exercises~\cite{kim2019crowdsourcing}, or responding to reflective prompts~\cite{zhang2021nudge}. While these studies demonstrate that text-based reflective nudges can positively influence the quality of individual contributions and, by extension, the overall deliberative process, they do not consider the potential of multimodal approaches to support diverse reflection styles. 

Given that individuals process information through various sensory channels~\cite{mayer2005cambridge}, integrating multimodal reflection mechanisms could further enrich online deliberation experiences~\cite{mayer2005cambridge}. While the role of multimodality in learning is well established in academic contexts (e.g.,~\cite{mayer2005cambridge}), its potential for fostering reflection in deliberative settings has not been as widely explored. The theoretical bridge between multimodal learning and reflection lies in the shared cognitive mechanisms that underpin both processes. This cognitive overlap suggests that reflection, much like multimodal learning, can be deepened by engaging multiple sensory channels~\cite{moon2013handbook}. We thus propose that individuals have distinct \textit{reflecting styles} akin to their learning styles, which can be effectively supported through multimodal reflective nudges. To our knowledge, no prior work has systematically explored how multimodal reflective nudges can improve deliberative outcomes.

To this end, we seek to address this gap by examining the impacts of different modalities on the quality of deliberation. Broadly, we ask the question, \textit{``How does the modality of reflection nudge affects deliberation quality?''} Our investigation involves designing and implementing an interface intended to induce self-reflection. Informed by literature~\cite{mayer2005cambridge, fadel2008multimodal, paivio1990mental, fleming2001vark, fleming1992not}, we focus on four different modalities (text, image, video, and audio) and applied them to two types of reflective nudges (Direct: Persona and Indirect: Storytelling) to facilitate their integration into existing online deliberation platforms. To support our examination in this aspect, we pose the following research questions: \textbf{RQ1: What are the preferred modalities for each of the reflective nudges?} and \textbf{RQ2: How does the modality of a reflective nudge affect the quality of deliberation?}

We conducted two user studies to address each question. The first study, involving 20 participants, explored the preferred modalities for reflective nudges using an early prototype of our system, powered by a Large Language Model (LLM) (i.e., GPT-4.0) and Generative AI. Participants engaged with reflective prompts in text, image, video, and audio formats, and we gathered insights on their preferences and comfort levels. Notably, text was the most preferred modality for direct reflective nudges (persona), while video was preferred for indirect reflective nudges (storytelling). The second study, an experimental investigation involving 200 participants, evaluated the impact of different modalities on deliberation quality. We developed eight interfaces, each integrating a distinct modality with a reflective nudge, and measured their impacts on deliberative outcomes. Our results highlighted that subjectively preferred modalities do not always produce higher levels of deliberativeness. Specifically, video leads to higher level of deliberative quality for persona. Our results also revealed that video worked well with both reflective nudges. We discuss later how different nudge types may have an \textit{`optimal'} modality, and how modality influences the dynamics of online deliberation.

We make the following contributions:
\begin{itemize}
    \item The design and integration of multimodal self-reflection interface nudges into online deliberation platforms, using LLMs for adaptability and diversification.
    \item An in-depth analysis of the modalities, highlighting the nuanced ways in which different modalities influence the quality and depth of discourse in online deliberation spaces.
\end{itemize}
\section{Background and Related Work} 
We begin by providing a background of our primary focus: internal reflection within the deliberation process and assessing deliberative quality. We then examine the role of modalities in supporting self-reflection, and their interaction with dual-system thinking. Finally, we review recent advancements in nudging techniques to enhance deliberative outcomes, along with the integration of reflection mechanisms with large language models, emphasizing the ethical implications of using AI in deliberation.

\subsection{Reflection in Deliberation}
\label{sec: reflection background}
Deliberation is a process that entails the careful and thoughtful consideration of diverse perspectives~\cite{davies2013online, goodin2003does, aristotle1984complete, hobbes1946}, allowing individuals to weigh reasons for and against a given measure~\cite{goodin2003does, oxford} before making informed choices grounded in thorough consideration and reasoned judgment~\cite{christiano2009debates}. Theorists of deliberative democracy emphasize that internal reflection is a foundational element of the deliberation process~\cite{muradova2021seeing, dryzek2002deliberative, goodin2000democratic, chambers2003deliberative}, preceding any public discussion~\cite{goodin2003does}. They assert that `\textit{deliberation is reflective rather than simply reactive}'~\cite{bohman2000public, dahlberg2001computer, janssen2004online}, requiring the collection of diverse information and development of a nuanced understanding of multiple perspectives, rather than an immediate response to an issue~\cite{davies2009online}. This pre-discussion, internal introspective phase is thus essential for individuals to clarify and solidify their stance, enabling more thoughtful contributions in subsequent discussions~\cite{goodin2003does, holdo2020meta}. 

Dryzek~\cite{dryzek2009democratization} proposed reflection as a metric to evaluate the reflective capacity of deliberative systems. Such a metric considers individuals' ability to incorporate others' perspectives and demonstrate public-mindedness and sincerity in their deliberations~\cite{holdo2020meta}. An early example is PICOLA (Public Informed Citizen Online Assembly)~\cite{cavalier2009deliberative} which comes with a reflection phase where participants could think about issues, and then discuss in an asynchronous forum. The results support an online conversation that is informed and structured, though it was not formally published~\cite{cavalier2009deliberative}. Similarly, prior research has explored novel interfaces that incorporate various reflective approaches to enhance deliberation~\cite{arceneaux2017taming}. For example, in Reflect~\cite{kriplean2012you}, an interface prompting users to restate points made by other commenters (e.g., ``\textit{What do you hear James [a commenter] saying?}''), fosters mutual understanding through clarification and dialogue. The same authors also extended Franklin's pro/con technique~\cite{franklin1956mr} for internal reflection to encourage participants to organize their reasoning into a list of pros and cons (i.e., ``\textit{Give your Cons. Give your Pros.}''~\cite{considerConsideritOnly, kriplean2012supporting, kriplean2011considerit}). This has helped participants to identify unexpected common ground. In another approach, \textbf{perspective-taking} is used to consider the views of various stakeholders~\cite{tuller2015seeing}. Kim et al.\cite{kim2019crowdsourcing} employed this via textual open-ended questions, to ``\textit{Guess the perspective of some other stakeholder groups (at least two)}.'' Similarly, the \textbf{imagined other} reflection approach~\cite{batson1997perspective, galinsky2000perspective} prompts participants to construct narratives from the perspective of an imagined person. Zhang et al.~\cite{zhang2021nudge} observed how \textbf{reflective questions} like ``\textit{What are your opinions on [an issue]?}'' and ``\textit{How might others disagree with you on [the same issue]?}'' improve attitude clarity and correctness, and encourage opinion expression. Additionally, Price et al.~\cite{price2002does} examined how \textbf{reflective prompts} (e.g. ``\textit{articulate the reasoning behind others' opinions}''), foster deeper deliberativeness. More recently, Yeo et al.~\cite{yeo2024help} found that textual reflective nudges, such as the textual display of other stakeholders' opinions or posing open-ended questions (e.g., ``\textit{How has your perspective on the impact of [an issue] changed as you have grown?}'') have significant positive effects on deliberativeness and can shape the dynamics of online discussions. Collectively, these empirical studies support two conclusions: (1) reflection approaches influence deliberativeness, a normatively desired effect for deliberation~\cite{bohman2000public, zhang2021nudge} and (2) most reflection approaches rely on the conventional text-based formats.

While the above approaches have effectively broaden the scope of reflection, they remain limited to text-based formats. This reliance overlooks potential benefits of a multimodal approach that could accommodate different cognitive and reflective styles~\cite{mayer2005cambridge}. In this study, we thus focus on examining the efficacy of different modalities in reflective approaches. By building on these foundational approaches, we expand the toolkit for facilitating reflection in deliberation and contributes to a broader understanding of the effects of distinct modalities on the deliberative process. 

\subsection{Assessing Deliberative Quality: Deliberativeness}
\label{sec: measurements}
Deliberativeness refers to the quality of deliberation made by an individual~\cite{trenel2004measuring}, a cornerstone of effective public discussions~\cite{menon2020nudge}. The term has been defined in diverse ways in the literature~\cite{graham2003search, price2002does, steenbergen2003measuring, stromer2007measuring}, with some papers~\cite{bohman2000public, price2002does, zhang2021nudge} employing the term `\textit{opinion quality}' interchangeably. In our study, we define deliberativeness as the quality of an individual's opinion~\cite{menon2020nudge}.

As a multifaceted construct~\cite{menon2020nudge, zhang2021nudge}, deliberativeness encompasses several dimensions~\cite{trenel2004measuring, price2002does, steenbergen2003measuring, stromer2007measuring}, with \textit{rationality} and \textit{constructiveness} forming the core conditions of deliberation~\cite{trenel2004measuring}. Rationality includes both \textit{opinion expression} and \textit{justification level}. Opinion expression evaluates the clarity and presence of articulated opinions necessary for meaningful deliberation~\cite{trenel2004measuring}. Originally developed for group contexts, opinion expression underpins rational deliberation~\cite{beck2000argumentation}, playing a pivotal role in shaping the effectiveness and productivity of subsequent discussions and decision-making~\cite{beck2000argumentation}. Bales~\cite{bales1983overview} connects opinion expression to the requirement for logical, well-reasoned arguments, while Fietkau and Trénel~\cite{fietkau2002interaktionsmuster} emphasize that clearly articulated opinions ensures structured, objective discussions by fostering clarity and substantive contributions. 
Justification level assesses the extent of \textit{reasoning} provided to support one's opinion~\cite{trenel2004measuring}. The ability to \textit{reason} is a recurring theme in Dryzek's work~\cite{dryzek2002deliberative}, where he argues that deliberation hinges on the act of reasoning, with the power of deliberative democracy lying in the ability to reason --- what Habermas~\cite{habermas2015between} famously described as ``\textit{the force of the better argument}''. 

Constructiveness, on the other hand, reflects the degree to which an individual integrates diverse perspectives and seeks common ground~\cite{goddard2023textual, trenel2004measuring}, moving beyond purely self-interested arguments~\cite{kymlicka2002contemporary}. It is evaluated based on whether opinions incorporate an inclusive consideration of varied perspectives or remain narrow and one-dimensional~\cite{trenel2004measuring, black2011self, gastil2007public}. 
Philosophical traditions, such as those of Plato and Aristotle~\cite{walton1999one}, and Garver~\cite{garver2004sake}, advocate the value of balanced, two-sided arguments for constructive deliberation, fostering understanding and cooperation. Similarly, Hackett and Zhao~\cite{hackett1998sustaining} argue that constructiveness involves presenting opposing views fairly to encourage diverse perspectives. Gastil~\cite{gastil2008political} highlights the importance of equitable argumentation in democratic deliberation, advocating for a mix of viewpoints. Niculae and Niculescu-Mizil~\cite{niculae2016conversational} found that constructive contributions are often more balanced, aligning with Parkinson and Mansbridge~\cite{parkinson2012deliberative}, who stress that inclusive discourse enhances deliberative outcomes.
Together, rationality and constructiveness serve as the core conditions for effective deliberation, ensuring both reasoned and inclusive arguments.

In addition to rationality and constructiveness, two other essential dimensions --- argument repertoire and argument diversity --- are integral to this definition, as they directly influence the quality of deliberation. \textit{Argument repertoire} measures the quality of opinions by counting non-redundant arguments presented both for and against an issue --- an approach widely adopted in deliberative research~\cite{menon2020nudge, zhang2021nudge, cappella2002argument, kim2021starrythoughts, yeo2024help}. Meanwhile \textit{argument diversity} measures the range of perspectives within an opinion. It reflects the inclusion of multiple perspectives and interpretations within an individual's opinion~\cite{anderson2016all, gao2023coaicoder, richards2018practical}, highlighting the breadth of viewpoints considered. This measure also aligns with constructiveness, as a broad range of arguments often demonstrates an openness to differing perspectives and a willingness to engage with opposing views. 

Building on prior work, deliberativeness in this study is assessed through the interplay of these five measures: rationality (comprising opinion expression and justification level), constructiveness, argument repertoire, and argument diversity. These measures collectively capture the depth, inclusivity, and quality of opinions within the deliberative process.

\subsection{Modalities, Learning and Self-Reflection}
\label{sec: learning and reflection}
Research on multimodality (i.e., the use of multiple content representations such as text, video, audio, images and interactive elements~\cite{sankey2010engaging}) has extensively demonstrated its effectiveness within learning environments~\cite{mayer2005cambridge, birch2005students, williamson2010learning, sankey2009ethics, sankey2010engaging, pashler2008learning, miller2001learning, mestre2010matching, picciano2009blending, sankey2011impact, dolzhich2019multimodality, gargallo2018perceptual, borzello2018benefits, schwartz2014s}, with consistent findings highlighting its benefits for information processing~\cite{bartholomeus1972acquisition, birch1964auditory, lawton1973developmental}, enhancing comprehension and problem solving~\cite{sankey2010engaging, schroeder2013effective, fadel2008multimodal}, as well as cognitive outcomes~\cite{constantinidou2002stimulus}. Key frameworks like Dual Coding Theory~\cite{paivio1990mental, bodemer2002encouraging, mayer2002multimedia, sweller1988cognitive} explain that combining visual and verbal inputs facilitates better encoding and retrieval by reducing cognitive load. This aligns with Moreno and Mayer~\cite{moreno1999cognitive}, who found that mixed-modality presentations yield better outcomes by leveraging the simultaneous processing of auditory and visual inputs in working memory~\cite{paivio2013imagery}. Mayer’s Cognitive Theory of Multimedia Learning~\cite{mayer2005cambridge} underscores the ``\textit{multimedia principle}'', showing that presenting information through words and images is more effective than text alone. Neuroscience research further supports these findings, revealing significant learning gains through integrated visual and verbal strategies~\cite{fadel2008multimodal}. Findings in the field of cognitive science suggest that ``\textit{intelligences and mental abilities exist along a continuum, responding to and learning from the external environment and instructional stimuli}''~\cite{gardner2011frames}, suggesting a framework for a multimodal instructional design to cater to varying cognitive processes and learning preferences~\cite{picciano2009blending}. These insights highlighted that multimodal learning not only encourages learners to develop versatile learning strategies~\cite{hazari2004applying, picciano2009blending} but also demonstrates that engaging multiple senses leads to more effective learning outcomes~\cite{kearsley2000online}.

Learning style theory further explores how individual preferences for modalities influence learning outcomes. Recent research indicates that tailoring educational strategies to these styles can significantly impact academic success~\cite{newble1985learning}. %Early studies, such as Galton's investigation on visual and verbal thinkers~\cite{galton1883inquiries}, suggest that individuals naturally select preferred modalities to interact with their environment. Bartlett~\cite{bartlett1995remembering} concluded that while individuals can adapt to various cues --- visual, auditory or kinesthetic --- they tend to favor one modality if left unprompted. This preference is often guided by selective filtering~\cite{bissell1971sensory, bruininks1969auditory}, where individuals prioritize certain modality inputs over others. 
A framework that is widely adopted for understanding modality selection in learning is the VARK model~\cite{fleming2001vark}, which categorizes learners into Visual (V), Auditory (A), Reading/Writing (R) and Kinesthetic (K) types~\cite{fleming1992not, fleming1995m, fleming2001vark}. %An extension of this is the ``\textit{meshing hypothesis}''~\cite{pashler2008learning, rogowsky2015matching}, which suggests that aligning instruction with a learner's preferred style improves learning outcomes. 
While empirical support for the framework is limited~\cite{pashler2008learning, rogowsky2015matching}, studies like Daniel and Tacker\cite{daniel1974preferred} and Waters~\cite{waters1972analysis} reported modest benefits when instructional methods matched modality preferences. Despite critiques of the VARK model%and the meshing hypothesis~\cite{pashler2008learning}
, VARK remains popular due to its practical applications in learning~\cite{sulistyanto2023effectiveness, noor2023bridging, el2024influence, laxman2014exploration} and its ability to foster multimodal engagement~\cite{sulistyanto2023effectiveness}. Aligned with VARK, the Dunn and Dunn Learning-Style Model~\cite{dunn1993teaching, dunn1984learning} categorizes learning styles into five stimuli: environmental, sociological, emotional, physiological (similar to VARK), and psychological. Meta-analyses~\cite{lovelace2005meta, dunn1995meta}, specifically one with 36 experimental studies~\cite{dunn1995meta}, alongside additional research~\cite{oweini2016effects}, affirm that tailoring instruction to these styles lead to significant gains in academic performance and learner satisfaction. These findings highlight that multimodal engagement enhances academic outcomes and fosters greater adaptability, reinforcing the value of personalized learning strategies.

While the role of multimodality in learning is well-documented, its broader implications, particularly in fostering reflection has received comparatively less attention. Since reflection is central to learning~\cite{kolb2014experiential}, it stands to reason that reflection may also benefit from a multimodal approach. Reflection is dynamic, involving deep engagement with diverse inputs, much like how learners engage with multimedia to absorb content~\cite{moon2013handbook}. Educational psychology studies also support that multimodal approaches can scaffold reflective processes, fostering critical and holistic thinking~\cite{fleming2001vark, mayer2005cambridge}.

For this study, we thus seek to assess how multimodal inputs (such as text, audio, images and video) influence reflective processes and their impacts on deliberative outcomes. By integrating insights from cognitive and educational psychology, we aim to explore how multimodality fosters deeper engagement and more effective reflection.

\subsection{Modalities and Dual-System Thinking}
\label{sec: dual system thinking}
Dual-system theories of thinking and decision-making classify cognitive processes into two primary systems: System 1, which is intuitive, fast, automatic, and emotional, operating unconsciously, and System 2, which is reflective, deliberate, analytical and effortful~\cite{kahneman2011thinking, strack2004reflective, sloman1996empirical}. %System 1 is the principal mode of thinking, dominating routine, quick decisions that require minimal effort such as walking or driving, while System 2 is engaged for complex reasoning that requires conscious effort~\cite{kahneman2011thinking}. Although System 2 supports critical thinking, System 1 excels in speed and multitasking~\cite{zavolokina2024think}, handling about 95\% of daily decisions through \textit{heuristics} --- mental shortcuts that simplify judgments~\cite{bargh2001automated}. 
Due to our predisposition to reduce effort, System 1
%these heuristics 
enables efficient decision-making% by using readily available information, often 
, yielding relatively accurate judgments~\cite{shah2008heuristics}, but
%while conserving cognitive effort and bypassing the need for deliberate reflection~\cite{shah2008heuristics}. However, relying on heuristics can 
exposes us to cognitive biases --- systematic errors that skew judgments from rationality~\cite{caraban201923, kahneman1991anomalies}. % For example, the status-quo bias causes individuals to favor default options over alternatives, even when those defaults are suboptimal~\cite{kahneman1991anomalies}. Sloman~\cite{sloman1996empirical} underscores the interplay between these two systems, emphasizing their complementary roles. Together, the two systems balance efficiency and critical analysis, enabling humans to navigate reasoning tasks that require both intuition and logic~\cite{kahneman2011thinking, sloman1996empirical}. 
 Consequently, we often need to explicitly engage System 2 for careful, rational deliberation~\cite{kahneman2011thinking, sloman1996empirical}.

In our study, we employ the dual-system theories as a framework to examine how different modalities support System 1 or System 2 thinking. Guided by Kahneman’s six distinct characteristics of dual-system thinking~\cite{kahneman2011thinking, kahneman2002maps, kahneman2012two} (see Table~\ref{tab: dual system thinking}), we adopt these characteristics to analyze how different modalities interact and best facilitate the two cognitive systems, helping us identify which best facilitate intuitive (System 1) or reflective (System 2) thinking.

\begin{table*}[!htbp]
\caption{Six characteristics of Dual-System Thinking based on~\cite{kahneman2011thinking, kahneman2002maps, kahneman2012two}. The sections highlighted in blue illustrate how these characteristics apply in our study to understand how different modalities interact with the two cognitive systems.}
\label{tab: dual system thinking}
\scalebox{0.77}{
\begin{tabular}{|c|l|c|c|}
\hline
\textbf{Characteristic} &
  \textbf{Definition} &
  \textbf{System 1} &
  \textbf{System 2} \\ \hline
Speed &
  \begin{tabular}[c]{@{}l@{}}Speed of thinking\\ {\color[HTML]{3166FF} Ability of the modality to allow users to process information quickly (System 1) or slowly (System 2).}\end{tabular} &
  Fast &
  Slow \\ \hline
Processing &
  \begin{tabular}[c]{@{}l@{}}Approach to handling thinking tasks\\ {\color[HTML]{3166FF} Ability of the modality to influence whether users engage in parallel processing, focusing on the big picture} \\ {\color[HTML]{3166FF} and broader concepts (System 1), or in serial processing, emphasising detailed, step-by-step analysis (System 2).}\end{tabular} &
  Parallel &
  Serial \\ \hline
Control &
  \begin{tabular}[c]{@{}l@{}}Degree of conscious oversight\\ {\color[HTML]{3166FF} Ability of the modality to enable users to engage in automatic, intuitive reflection (System 1) or deliberate, more} \\ {\color[HTML]{3166FF} conscious control over the reflection process (System 2).}\end{tabular} &
  Automatic &
  Controlled \\ \hline
Effort &
  \begin{tabular}[c]{@{}l@{}}Cognitive load\\ {\color[HTML]{3166FF} Ability of the modality to engage users with varying levels of cognitive effort, ranging from low mental effort (System 1)} \\ {\color[HTML]{3166FF} to high mental effort and concentration (System 2).}\end{tabular} &
  Effortless &
  Effortful \\ \hline
Nature &
  \begin{tabular}[c]{@{}l@{}}Inherent operating mechanism\\ {\color[HTML]{3166FF} Ability of the modality to align with users' inherent reflecting styles and guide them toward either intuitive thinking} \\ {\color[HTML]{3166FF} (System 1) or rule-governed, analytical reasoning (System 2).}\end{tabular} &
  Associative &
  Rule-governed \\ \hline
Adaptability &
  \begin{tabular}[c]{@{}l@{}}Ability to change or evolve\\ {\color[HTML]{3166FF} Ability of the modality to accommodate users' needs by adapting to the complexity of topics and varying levels} \\ {\color[HTML]{3166FF} of prior knowledge, allowing for either slow-learning, rigid thinking (System 1) or flexible, adaptive reasoning (System 2).} \end{tabular} &
  Slow-learning &
  Flexible \\ \hline
\end{tabular}}
\Description{The table provides a summary of the six characteristics of dual-system thinking: speed, processing, control, effort, nature, and adaptability. Each characteristic is defined, and the table illustrates how these characteristics are applied within the context of our study. Additionally, the table shows a comparison between System 1 (intuitive, fast, and automatic thinking) and System 2 (deliberative, slow, and controlled thinking) to explain the varying cognitive approaches influenced by different modalities.}
\end{table*}

\subsection{Nudging}
\label{sec: nudging}
Thaler and Sunstein~\cite{thaler2008nudge, Thaler_2009} introduced the notion of nudging to guide optimal decision-making by leveraging systematic biases in reasoning. A nudge is any modification to the choice architecture that predictably influences behavior without restricting options or altering incentives significantly~\cite{caraban201923}. Nudges work by making certain choices easier or more accessible, subtly steering individuals toward better decisions~\cite{thaler2008nudge}. By altering how choices are presented, nudges effectively intervene and redirect individuals away from their habitual modes of thought~\cite{thaler2008nudge, Thaler_2009}.% For instance, switching organ donation policies from opt-in to opt-out significantly enhanced societal welfare~\cite{caraban201923}, while strategically placing fruits near checkout counters increased healthier purchases, even when less healthier options such as cake remained available~\cite{Thaler_2009}. Both approaches preserved individual freedom of choice while subtly guiding behavior~\cite{thaler2008nudge}.

The idea of nudging has been applied across various domains, including HCI~\cite{caraban201923, lee2011mining}. Caraban et al.\cite{caraban201923} reviewed 71 HCI studies, identifying 23 distinct nudging mechanisms, categorized into six groups. Hansen and Jespersen~\cite{hansen2013nudge} further categorized nudges into four types based on two factors: the mode of thinking (automatic or reflective which is linked to the dual-system theories; see section~\ref{sec: dual system thinking}) and transparency of the nudge (whether the user perceives the nudge's intent). Nudges in the form of simple interface changes have also shown significant effects, such as Harbach et al.'s~\cite{harbach2014using} redesign of Google Play Store's permissions dialogue to encourage users to consider privacy risks in apps, as well as Wang et al.'s~\cite{wang2014field} use of visual cues and timers to reduce regret posting in online disclosures.

Within the realm of deliberation, nudges involving simple interface changes have proven effective in enhancing deliberativeness~\cite{xiao2015design, zhang2013structural}. Murray et al.~\cite{murray2013supporting} designed three reflective tools to enhance quality deliberations by fostering what they called ``social deliberative skills'' like perspective-taking and reflecting on one's biases. Their findings revealed that simple designs of the tools acted as scaffolding for social deliberative skills, enhancing quality deliberations. In other deliberative studies, Menon et al.~\cite{menon2020nudge} showed that interface nudges like partitioned text fields increased reply length by 35\% and argument count by 25\%. Similarly, Zhang et al.~\cite{zhang2021nudge} explored reflective nudges through question prompts such as ``What are your opinions on this issue?'', and found that they improved opinion quality and enhanced opinion expression. These studies highlighted the impact of small interface adjustments on fostering higher quality deliberation.

Drawing from existing efforts, we utilize the concept of nudges to improve the overall deliberativeness of online discourse. For this specific case, we examine how different modalities enacted on interface-based reflective nudges influence various dimensions of deliberative quality, anticipating that they enhance deliberativeness.

\subsection{Leveraging Generative AI for Reflection: Opportunities and Ethical Considerations}
The widespread adoption of information and communication technologies (ICTs)~\cite{janssen2018innovating}, including AI such as large language models (LLMs)~\cite{duberry2022artificial}, has significantly advanced deliberation and citizen-government relations~\cite{duberry2022artificial, lovejoy2012information, misuraca2020ai, mehr2017artificial, chun2012social, leach2010dynamic}. In reflective deliberation, LLMs' ability to generate personalized responses offers unique opportunities for fostering reflective thought at scale~\cite{duberry2022artificial, fogg2002persuasive, jakesch2023co}. While their use in this context remains nascent, early work by Yeo et al.~\cite{yeo2024help} demonstrated the potential of LLMs for broadening perspectives through textual reflection. However, the role of non-textual modalities has yet to be explored. In this study, we extend this work by integrating GPT-4.0 with generative AI for multimedia modalities to stimulate reflection and support deliberation on societal issues.

Despite these opportunities, LLMs are accompanied by risks. As LLMs become embedded in human communication~\cite{bommasani2021opportunities}, generating human-like language for widely-used applications like writing support~\cite{dang2022beyond, jakesch2023co, hancock2020ai} and grammar correction~\cite{koltovskaia2020student}, they influence opinions through a process termed \textit{latent persuasion}~\cite{jakesch2023co}. Similar to nudging (see section~\ref{sec: nudging}), latent persuasion subtly shapes decision-making by leveraging machine-generated language, underscoring AI's profound impact on thought processes~\cite{fogg2002persuasive, leonard2008richard}. With AI's growing prevalence, ethical concerns about its potential to amplify societal biases are rising~\cite{stahl2016ethics}. LLMs trained on biased language patterns~\cite{caliskan2017semantics, bolukbasi2016man, duberry2022artificial} risk perpetuating stereotypes and societal prejudices~\cite{brown2020language, lucy2021gender, huang2019reducing, nozza2021honest, johnson2022ghost, langer2023trust, bowen2006information, labajova2023state, cotter2019playing}, shaping user perceptions and reinforcing inequalities. The rise of deepfake technology, which creates realistic yet fabricated media, further complicates these concerns~\cite{karnouskos2020artificial, verdoliva2020media}.

Our study does not focus on using generative AI to verify the veracity of any content but instead as a tool to stimulate reflection. While addressing these inherent risks is beyond the scope of this work, it is still crucial to ensure ethical and responsible AI use. To mitigate the risks of stereotyping and biases in AI-generated content, our study employs carefully designed prompts (detailed in section~\ref{sec: section3} and Appendix Tables~\ref{tab: prompt engineering} and~\ref{tab: prompt constraint rationale}) to encourage open, inclusive reflections. For every reflective nudge, we generate multiple stakeholders, to make sure that multiple sides are represented during the reflection process. Doing so, we aim to promote deeper engagement and broader perspective-taking, fostering nuanced deliberation without influencing conclusions or imposing definitive answers. 
\section{Choices of the Nudges and Modalities}
\label{sec: section3}
We now outline the design of our interface nudges and the rationale behind the selection of both the nudges and the modalities.

\subsection{Choice of Nudges}
Our goal was to consider nudges that would trigger different types of self-reflection. We identified two types: \textbf{direct reflective nudges} and \textbf{indirect reflective nudges}. \textbf{Direct} reflective nudges typically involve explicit, targeted prompts that encourage individuals to reflect on their own thoughts, experiences or opinions, and are usually shorter, while \textbf{indirect} ones are subtler, often encouraging reflection through external examples or narratives, allowing users to reflect by considering others' viewpoints or scenarios and usually feature more content. This approach allows us to explore how different modalities support varying levels of reflective nudges.

\subsubsection{Direct Reflective Nudge: Persona}
The direct reflective nudge employs a \textit{persona} approach, encouraging reflection through perspective-taking~\cite{galinsky2000perspective, kim2019crowdsourcing, zhang2021nudge}. This approach is grounded in principles from constructivist learning theories~\cite{vygotsky1978mind, piaget1985equilibration}. By guiding users to explicitly examine their own thoughts, the persona approach mirrors Vygotsky's concept of mediated learning~\cite{vygotsky1978mind}, enabling users to internalize insights through active interaction with reflective stimuli. Piaget's theory~\cite{piaget1985equilibration} complements this by focusing on how learners construct knowledge through self-directed exploration. Reflection, as facilitated by the persona, prompts users to critically evaluate their preconceptions, fostering new understandings and alignment with Piaget’s notion of active knowledge construction. Moreover, we specifically chose persona as its ability to reach higher levels of deliberativeness compared to others, have been established in previous work~\cite{yeo2024help}.

\subsubsection{Indirect Reflective Nudge: Storytelling}
The indirect reflective nudge employs a \textit{storytelling} approach, encouraging users to reflect through narratives that present others' perspectives~\cite{batson1997perspective, yeo2024help}. By fostering reflection through vicarious experiences, storytelling provides richer contextual information than the persona-based approach by extending its content with an added storyline. Grounded in narrative-based learning~\cite{bruner1991narrative, green2000role}, storytelling is described by Bruner~\cite{bruner1991narrative} as a fundamental mode of thought, enabling the construction of personal and social meaning through interpretation. Green and Brock~\cite{green2000role} further explore the concept of `transportation' --- the immersive mental absorption into a narrative, characterized by focused attention, emotional engagement and vivid imagery. Their findings highlight the persuasive power of storytelling in shaping beliefs and attitudes. Additionally, prior research demonstrates that storytelling improves deliberativeness by engaging emotional and cognitive processes, ultimately enhancing critical thinking and reflective engagement~\cite{yeo2024help}.

\paragraph{\normalfont{Both nudges leverage self-referential encoding, where individuals process and internalize information by relating it to their own life, enhancing engagement and cognitive retention~\cite{rogers1977self}. These nudges also align with Kahneman's dual-system theories~\cite{kahneman2011thinking} (see section~\ref{sec: dual system thinking}), with the direct persona nudge engaging System 2 (i.e., analytical, reflective thinking) and the indirect storytelling nudge activating System 1 (intuitive, empathetic processing). Together, these nudges provide a multifaceted approach to fostering reflection.}} 

\subsection{Choice and Design of Modalities}
\label{sec: Modalities}
The design of each modality was guided by principles from multimedia learning theory~\cite{mayer2005cambridge, fadel2008multimodal} to enhance reflective engagement. We chose a set of modalities representative of non-interactive multimedia formats identified in prior research~\cite{mayer2005cognitive, de2005multimedia, deimann2006volitional, fadel2008multimodal}, ensuring they were representative of diverse content representations. According to Fadel~\cite{fadel2008multimodal}, non-interactive multimodal learning includes combinations such as text with visuals, audio and video formats. Empirical studies~\cite{mayer2005cognitive, de2005multimedia, deimann2006volitional, fadel2008multimodal} demonstrate that non-interactive, multimodal learning significantly improves learning outcomes compared to traditional single-mode approaches. Interestingly, when these scenarios shift from non-interactive to interactive settings, these gains are not statistically significant. Guided by this evidence, we identified the following non-interactive modalities to support reflection:
\begin{enumerate}
    \item \textbf{Text}: A traditional and widely used medium for reflective prompts. It follows the design principles of Cooper for persona~\cite{cooper1999inmates, cooper2007face, pruitt2010persona} and Freidus et al.~\cite{freidus2002digital} and Bruner~\cite{bruner1991narrative} for storytelling. As text is the most conventional modality, it is structured to optimize engagement without overwhelming users.
    \item \textbf{Image}: This modality integrates visual stimuli to enhance reflection, featuring the same text content, with additional images. It follows the \textit{spatial contiguity principle}, ensuring that images are placed alongside related text to minimize split-attention and enhance understanding~\cite{hegarty1993constructing, ayres2005split, moreno1999cognitive, mayer1989systematic, mayer1995generative, chandler1992split}. 
    \item \textbf{Audio}: Designed to leverage auditory cues, this modality employs the \textit{voice principle}, which suggests that reflections are more effective when delivered in a natural, human voice with a standard accent rather than a machine or foreign-accented narration~\cite{atkinson2005fostering}. This conversational delivery ensures engagement and encourages deeper thought.    
    \item \textbf{Video}: Combining visual and auditory elements, video content leverages its dynamic nature to simulate real-world experiences. This modality adheres to the \textit{temporal contiguity principle}, synchronizing narration with animations to ensure that verbal and visual information are presented together for better cognitive processing~\cite{moreno1999cognitive, mayer1991animations, mayer1992instructive, mousavi1995reducing}. Additionally, Paivio’s \textit{Dual-Coding Theory} supports the use of auditory and visual channels to reinforce reflective thinking~\cite{paivio2013imagery}.
\end{enumerate}

\begin{figure*}[!htbp]
  \centering
  \includegraphics[width=\linewidth]{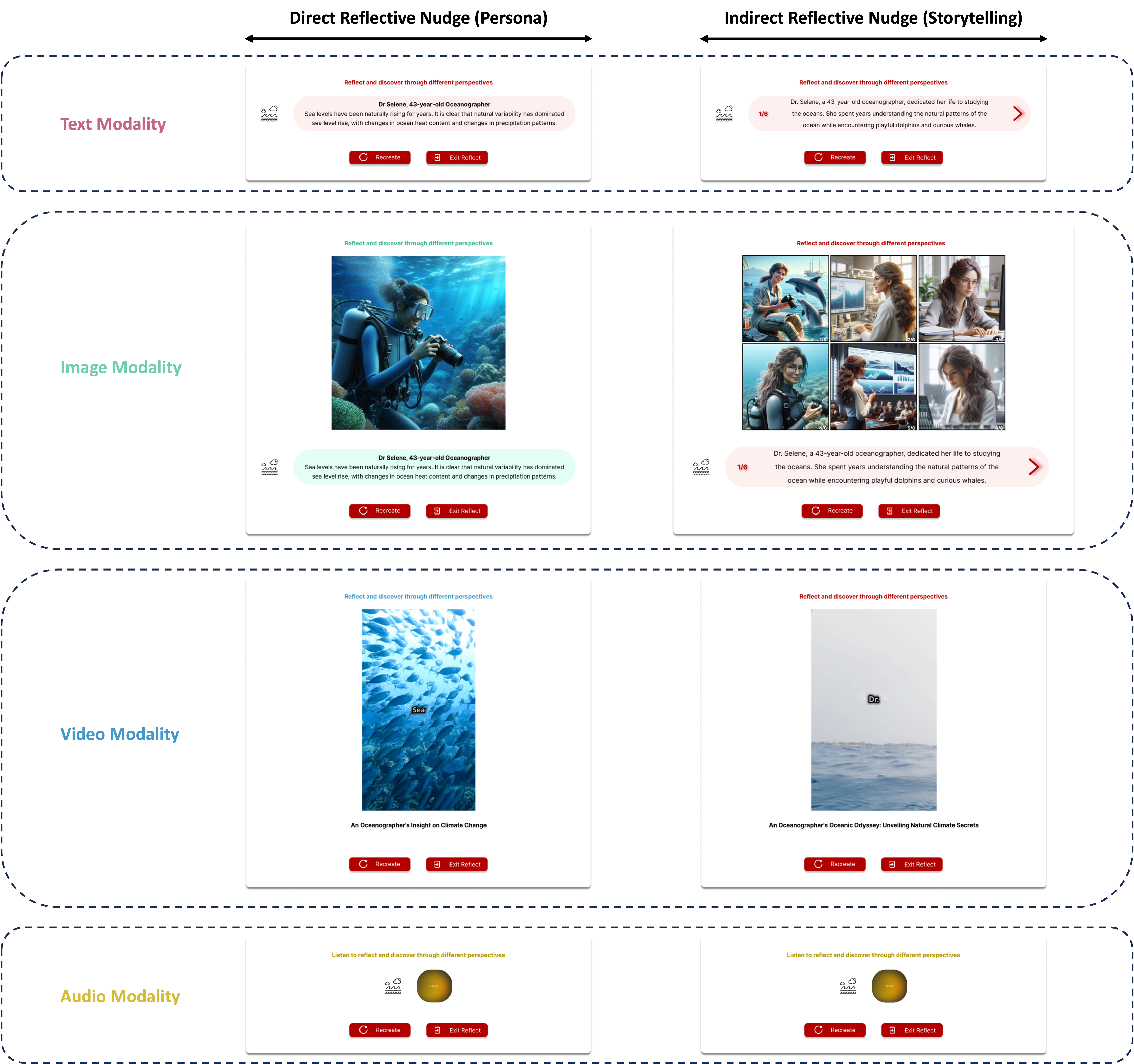}
  \caption{Example of the Two Reflective Nudges and their Respective Four Modalities.}
  \label{fig: modalities}
  \Description{Example of the two selected reflective nudges and their respective four modalities. An example for the text, persona: Dr Selene, 43 year old oceanographer. Sea levels have been naturally rising for years. It is clear that natural variability has dominated sea level rise, with changes in ocean heat content and changes in precipitation patterns. An example for the image, persona: an image of an oceanographer is shown with the exact same text shown at the bottom of the image. An example for the video, persona: a video is shown with the title - an oceanographer's insight on climate change. An example for the audio, persona: an audio is shown. An example for the text, storytelling: Dr Selene, a 43 year old oceanographer, dedicated her life to studying the oceans. She spent years understanding the natural patterns of the ocean while encountering playful dolphins and curious whales. An example for the image, storytelling: six different images were shown with the exact same text shown at the bottom of the images. Video and audio formats for storytelling is the same for persona except for its duration.} 
\end{figure*}

Figure~\ref{fig: modalities} provides examples for each of the modality and their respective reflective nudges. Content across modalities in each nudge type is the same.

\subsection{Implementation of the Modalities with LLM and Generative AI}
We used GPT-4.0 to generate a range of 10 textual prompts for the direct reflective nudge. We then instructed the LLM to generate short narratives for each of the same textual prompts for indirect reflective nudge. Following, to generate modalities, we used text-to-image generation tools (Bing Image Creator), text-to-video generation tool (Invideo AI) and text-to-speech generation tool (Narakeet AI). This ensures that the content across \textbf{all modalities is strictly the same}, just presented differently by their own modality, such that we can assess the impacts of the modalities on deliberativeness and not due to the quality of the content presented in each modality.

The utilization of LLMs to promote self-reflection on online deliberation platforms ensures the adaptability and scalability of reflective nudges.
%Using LLMs, we tap into its vast knowledge and language proficiency in enabling reflective nudges to accommodate to a diverse array of topics commonly found in online discussions. 
%This enhances and versatility of the reflective nudges to effectively cater to the varied and dynamic nature of discussions on online deliberation platforms.

\paragraph{\normalfont{\textbf{Generating textual variants for both nudges:} In creating the textual variants for the direct reflective nudge (persona), GPT was tasked with the role of ``\textit{a helpful assistant focusing on supporting users' self-reflection on a given topic}.'' We adhered to scholarly design principles for each reflective nudge, ensuring that the output aligned with the objective of each nudge. Following White et al.~\cite{white2023prompt}, we used a structured prompt template: defining the task, adding constraints, and setting clear expectations. This led us to prompt GPT with, ``\textit{Create ten distinct personas representing different perspectives on the topic. Provide the name, age, and occupation for each persona}.'' Specific constraints were also established: ``\textit{Create three male and three female personas}'' to mitigate gender bias, as prior research has shown that LLMs such as GPT-3 can reinforce gender stereotypes~\cite{brown2020language, lucy2021gender, huang2019reducing, nozza2021honest, johnson2022ghost}.}}

For the textual variants in indirect reflective nudge (storytelling), each persona generated above was then used as input to prompt GPT to create a unique story.

\paragraph{\normalfont{\textbf{Generating image variants for both nudges:}} To generate the image variants, we input GPT's text output into the text-to-image generator, prompting it with, ``\textit{Create a photo-realistic image with realistic textures and lighting of the [persona/story]}.'' For the indirect nudge (storytelling), we prompt the AI to generate individual images representing key moments in the story, which are then combined to create a cohesive visual narrative. A key constraint for images is to exclude any wordings, as AI-generated images often produce distorted or unreadable text~\cite{keyes2023hands}.}

\paragraph{\normalfont{\textbf{Generating video variants for both nudges:}}
For video, similarly, the text output from GPT was input into the text-to-video generator. We prompted it with, ``\textit{Create a photo-realistic video featuring the [persona/story]},'' specifying background details and voice narration. We constrained the model to adhere strictly to the provided script, ensuring that the content remained the same across all modalities.}

\paragraph{\normalfont{\textbf{Generating audio variants for both nudges:}}
For audio, likewise, the text output from GPT was input into the text-to-speech generator. We then manually selected a gender-appropriate voice that matches the gender specified in the text output. Audio prompts were not required for this process as the generator automatically converts the provided text into audio.}

All prompts, along with the rationale behind the constraints and generation tools for each modality, are detailed in Appendix Tables~\ref{tab: prompt engineering} and~\ref{tab: prompt constraint rationale}. The temperature parameter was set at 0.7 to maintain diversity in responses without introducing excessive randomness.
\section{Study 1: Subjective Modality Preferences}

To address \textbf{RQ1: What are the preferred modalities for each of the reflective nudges?} and discern the preferred modalities for each reflective nudges, we conducted a user study followed by semi-structured interviews with participants ($N=20$). The study focused on participants' preferences for specific modalities within each nudge type and examined how these modalities facilitated self-reflection in an online deliberation context. Detailed descriptions of the four modalities and the two reflective nudges are outlined in section \ref{sec: section3}. 

\subsection{Independent and Dependent Variables}
The study employed a 2 Reflective Nudge: \{Direct, Indirect\} $\times$ 4 Modality: \{Text, Image, Video, Audio\} mixed factorial design. \textit{Reflective Nudge} was between-subject, while \textit{Modalities} was within-subject. 
Within each nudge, the system could generate alternative content (variants), based on pre-recorded prompts. These would be the same for every modality.
The sequence of modality presentation and the order of variants within each modality were randomized for each participant to mitigate any potential ordering effects. 
We had one dependent variable, Ranking, which indicates users' preferences and was measured on a scale from 1 to 4 (for each modality). 

Results between nudges are not compared against each other; instead, the focus is on evaluating the effects of the different modalities within each nudge.

\subsection{Covariates}
\label{sec: Covariates}
%Following Yeo et al.~\cite{yeo2024help},
We controlled for four covariates as fixed factors to account for individual differences that could affect participants' engagement with the modalities.
%This inclusion helps ensure that the findings are not confounded by variability in these covariates. 
We included these covariates in the main analyses by using ANCOVA instead of ANOVA to control for their influence.

\paragraph{\normalfont{\textbf{Topic Knowledge and Topic Interest (TK-TI)}} was included to assess participants' familiarity and information access on the discussion topic. Prior research~\cite{zhang2021nudge} demonstrates that reflection interacts with information access to influence perceived issue knowledge. Mayer and Gallini~\cite{mayer1990illustration} also found that individuals with low prior knowledge benefited more from a combination of text and images. %Likewise, Kalyuga et al.~\cite{kalyuga1998levels, kalyuga2000incorporating} found that individuals with limited prior knowledge gained the most from integrated presentations, such as video. 
Whereas Moreno and Mayer~\cite{moreno1999cognitive} found that people with high prior knowledge are more likely to benefit from videos. TK-TI was evaluated through a five-item multiple choice questionnaire and two matrix tables comprising of 7-8 topic-related statements, and is measured by a score between 0 to 36.}
%(see Appendix Figure~\ref{}). Each multiple choice item was scored 1 for demonstrated knowledge or interest, while matrix statements were rated up to 3, with 0 otherwise, yielding a total score ranging from 0 to 36.

\paragraph{\normalfont{\textbf{Self-Reflection and Insight Scale (SRIS) questionnaire}} was included to account for individuals' inherent predisposition to reflect~\cite{grant2002self, silvia2022self}. It is a widely utilized self-reported scale, rooted in theories of meta-cognition and personal development~\cite{grant2001rethinking, grant2003impact}. It consists of 20 questions on a 1–6 Likert scale (1=strongly disagree, 6=strongly agree).} %to evaluate individual's self-reflective capacity and tendency for self-reflection. 

\paragraph{\normalfont{\textbf{Inherent Reflecting Styles.}}} We used the VARK version 8.02\footnote{https://vark-learn.com/the-vark-questionnaire/}, consisting of 16 question items. Scores for each modality category were subsequently computed to assess whether individuals have a clear preference for one modality over the others. 

\paragraph{\normalfont{\textbf{External Exposure and Interactions.}}} While reflecting styles account for individuals' inherent preferences, external exposure and interactions capture individuals' habitual engagement with various modalities across media platforms such as news articles and blogs for text; Pinterest and social media posts for images; TikTok, Instagram Reels, and YouTube Shorts for video; and podcasts for audio. Regular exposure to specific media formats may lead to stronger affinity for and preference towards those modalities~\cite{knobloch2014choice}. Participants reported their frequency of interaction with these different media platforms on a 1-5 Likert scale (1=Never, 2=Rarely (less than once a week), 3=Occasionally (1-3 times a week), 4=Frequently (4-6 times a week), 5=Daily).

\subsection{Apparatus}
The study was conducted remotely using Zoom. Participants were instructed to access a simulated online deliberation environment and to share their screen. This environment was based on an interface similar to Reddit~\cite{horne2017identifying, medvedev2019anatomy}, to leverage a familiar interface.

\subsubsection{Implementation}
The prototype was developed using Figma\footnote{https://www.figma.com/} before its deployment on Useberry\footnote{https://www.useberry.com/}, a user testing site. We developed the feature - \textit{Reflect} - which displays the four modalities of the reflective nudge.

\subsubsection{Key Features of the Prototype}
\begin{figure*}[!htbp]
  \centering
  \includegraphics[width=\textwidth]{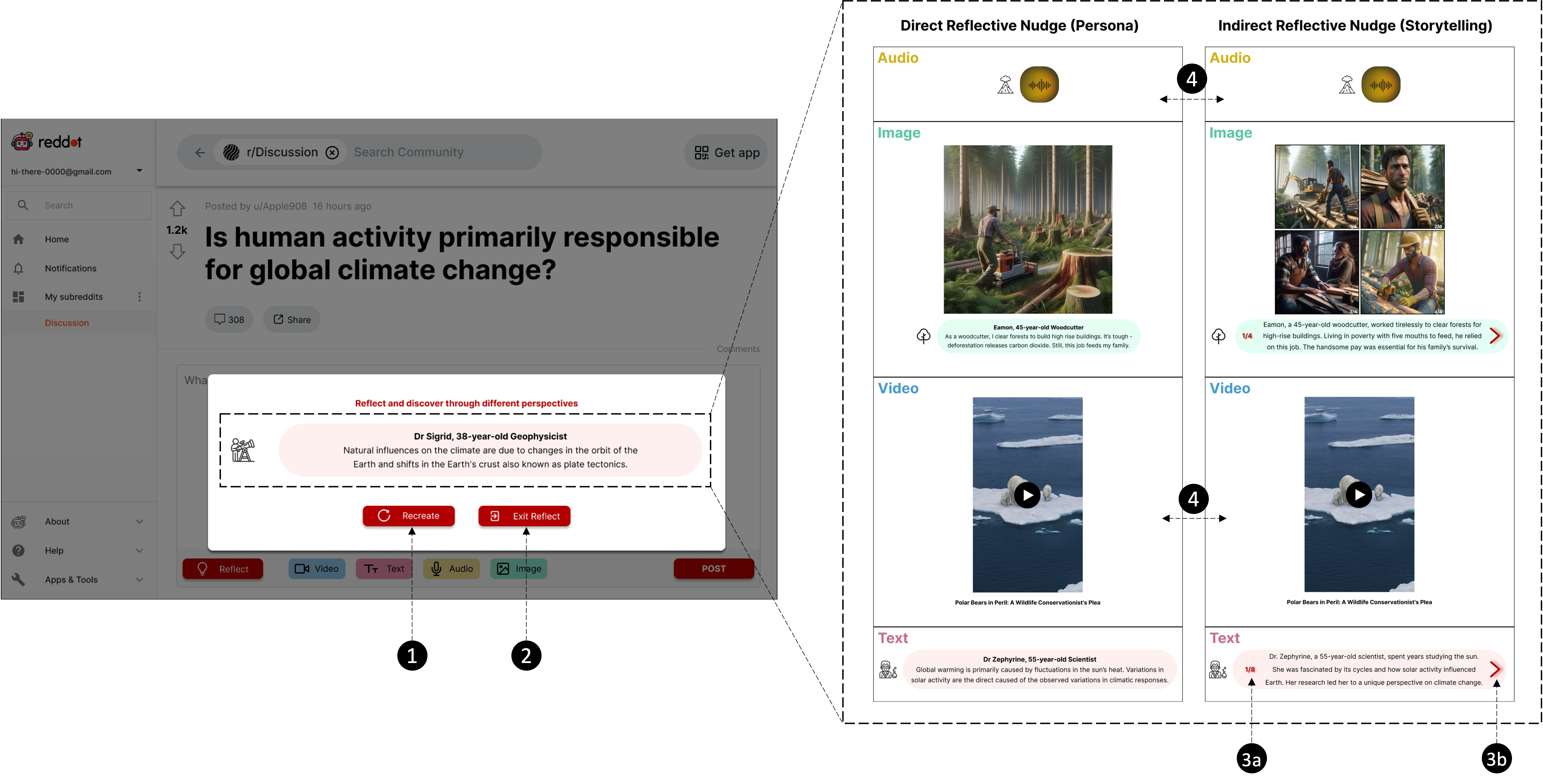}
  \caption{Key features in interface. \textbf{1:} Users can click on the \textit{Recreate} button to browse through different variants under a specific modality or \textbf{2:} click on the \textit{Exit Reflect} button to navigate back to the different array of modalities to choose another modality. Additional features are present for storytelling to allow users to \textbf{3a:} track their reading process and \textbf{3b:} continue with their reading on the current story. \textbf{4:} For audio and video formats in both direct and indirect reflective nudges, the primary difference is duration. In direct reflective nudges, audio and video average 9 to 12 seconds, while in indirect nudges, the duration extends to 90 to 120 seconds, reflecting typical short-form media content.}
  \label{fig: features}
  \Description{The user interface created for the study encompasses the following features in the comment box: 1: Users can click on the Recreate button to browse through different variants under a specific modality or 2: click on the Exit Reflect button to navigate back to the different array of modalities to choose another modality. When clicking on the storytelling reflector, additional features are present: 3a: The numbers, for example, page 1 out of 5 allows users to track their reading process. 3b: The forward arrow allows users to click and continue with their reading on the current story.}
\end{figure*}

To facilitate users' self-reflection during the writing process on a discussion topic, we designed several key features. These are depicted in Figures~\ref{fig: teaser} and \ref{fig: features}.

\paragraph{\normalfont{\textit{Reflect}} is a feature that can be integrated into online discussion platforms to facilitate users in their self-reflection process when they craft their opinions on a discussion topic. Clicking on \textit{Reflect} displays the four distinct modalities. Users can choose one to explore, allowing for a deeper engagement within that modality’s context. These are depicted in Figure~\ref{fig: teaser} (3a and 3b).} 

\paragraph{\normalfont{\textit{Recreate}} generates variants tailored to that modality (Figure~\ref{fig: features} (1)). To avoid undue one-sidedness in the generated content, we ensured a balanced presentation by alternating variants between male and female perspectives, as well as between positive and negative tones for both nudges.}

\paragraph{\normalfont{\textit{Exit Reflect}} will return to the selection screen displaying all available modalities (Figure~\ref{fig: features} (2)).}

\paragraph{Features Specific to the Indirect Reflective Nudge}
While engaging with a story, users can track their progress using the page tracker depicted in Figure~\ref{fig: features} (3a). This tracker displays the number of remaining pages until the story concludes. Stories vary in length: four (short), six (medium) or eight (long) pages. Clicking the \textbf{\huge{>}} button (Figure~\ref{fig: features} (3b)) allows users to continue reading. In direct reflective nudges, audio and video have a duration of 9 to 12 seconds, while in indirect nudges, it extends to 90 to 120 seconds, reflecting typical short-form media content.

\subsection{Participants and Ethics}
Following approval from our Institutional Review Board (IRB), we recruited 20 participants (8 males and 12 females), with 10 participants assigned to each type of reflective nudge, aligning with local sample size guidelines~\cite{caine2016local}. The participants had an average age of 24.0 years ($SD=3.56$). Notably, remote interviews typically have a mean sample size of 15 ($SD=6$), and CHI publications commonly feature 12 participants~\cite{caine2016local}. All participants were university students, with detailed demographic information for each reflective nudge provided in Appendix Table~\ref{tab: st1-demo}. Participants were compensated at the appropriate rate dictated by our local IRB.

\subsection{Task and Material}
\label{sec:maintask}
We chose the discussion topic “Is human activity primarily responsible for global climate change?” from ProCon.org\footnote{https://www.procon.org/} as a task for its accessibility and relevance, thereby promoting constructive and open debate.

In the main task, participants used our prototype with the \textit{Reflect} feature. \textbf{Participants had to write at least 30 words with the help of the four modalities when expressing their perspective on the discussion topic.} Participants engaged with all four modalities sequentially (as the study was conducted over Zoom, the researcher instructed and guaranteed this implementation). 

We wanted to capture an organic interaction with the system, therefore did not guide when to use \textit{Reflect}. Participants could concurrently write while using it, or use it sequentially before or after writing. Additionally, \textbf{participants were not obliged to explore every variant within a modality, though they had the option to do so if desired.}

\subsection{Procedure}
\label{sec: procedure}
In addition to the main study (see section~\ref{sec:maintask}), we included a pre- and post-task.

\subsubsection{Pre-Task: Consent, Instructions, Demographic, Questionnaire for Covariates}
Participation consent was obtained before the study. Participants were informed that they had to express their opinions on a discussion topic, which was undisclosed at this point, using various modalities. Participants were then randomly assigned to one of two groups: direct or indirect reflective nudges. Following this, they completed a demographic questionnaire.

Before starting, we administered four questionnaires corresponding to each of the four covariates, as detailed in section~\ref{sec: Covariates}. The values of the covariates are summarized in Appendix Table~\ref{tab: st1-demo}, with no significant differences observed between the two participant groups for any covariate.

\subsubsection{Post-Task Interview}
Following the completion of the main task, we conducted semi-structured interviews to collect feedback. Participants were asked to rank the four modalities on a scale from 1 (lowest) to 4 (highest) based on their personal preference and the level of self-reflection each modality elicited. They provided explanations for their rankings and discussed specific challenges encountered with each modality, as well as the overall usefulness of each modality in their reflective process. All interviews were audio recorded and transcribed for subsequent analysis.

Participants took an average of 46.8 minutes to complete the study.

\subsection{Findings - Ranking}
\begin{figure*}[!htbp]
  \centering
  \includegraphics[width=.8\textwidth]{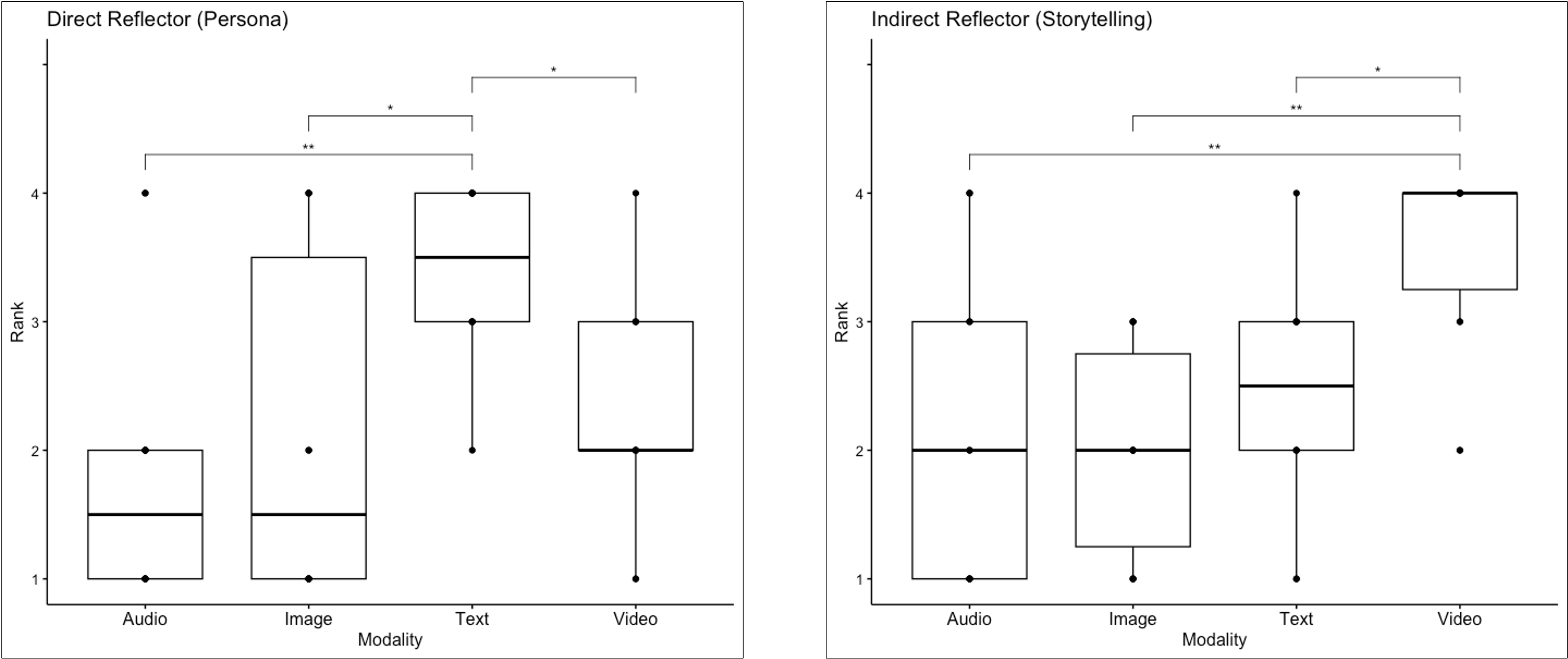}
  \caption{Mean ranking for the four modalities for the direct reflective nudge (left) and indirect reflective nudge (right) with 1 being the lowest rank and 4 being the highest rank. We report the results of the ANCOVA test and pairwise comparisons with BH correction, where * : p < .05, ** : p < .01.}
  \label{fig: ranking}
  \Description{Box plot describing the mean rankings for the four modalities for direct reflective nudge (left) and indirect reflective nudge (right). The X-axis shows the different modalities and the Y-axis shows the ranking value ranging from 0 to 4 at a rank interval of 1. 1 is the lowest possible rank and 4 is the highest possible rank. The highest mean ranking was found in text for direct reflective nudge at 3.40, while the highest mean ranking was found in video for indirect reflective nudge at 3.60.}
\end{figure*}

Figure~\ref{fig: ranking} shows the average ranking for each modality of the reflective nudges. 

\subsubsection{Direct Reflective Nudge (Persona)}
\emph{Text} received the highest average ranking of 3.4 out of 4, while \emph{Audio} received the lowest average ranking of 1.7 out of 4. A one-way repeated measures ANCOVA was conducted to determine a statistically significant difference between Modalities on Ranking. We found a statistically significant main effect of Modalities on Ranking ($F_{3,26} = 4.13$, $p<.05$). Specifically, \emph{Text} was significantly ranked higher compared to \emph{Audio} ($p<.01$), \emph{Image} (2.1 out of 4) and \emph{Video} (2.3 out of 4) (both $p<.05$).

\subsubsection{Indirect Reflective Nudge (Storytelling)} 
\emph{Video} received the highest average ranking of 3.6 out of 4, while \emph{Image} received the lowest average ranking of 2.0 out of 4. A one-way repeated measures ANCOVA revealed a statistically significant main effect of Modalities on Ranking ($F_{3,26} = 4.59$, $p<.01$). Specifically, \emph{Video} was ranked significantly higher than \emph{Audio} (2.2 out of 4) and \emph{Image} (both $p<.01$), and \emph{Text} (2.4 out of 4) ($p<.05$).

\paragraph{\normalfont{In summary, \emph{Text} was the most preferred modality for direct reflective nudges, while \emph{Video} was most favored for indirect reflective nudges.}}

\subsection{Findings - Subjective Feedback}
\label{sec: qualitative for study 1}
We present a detailed account of participants' experiences with each modality for each reflective nudge type, highlighting both benefits and challenges. Feedback is categorized using Kahneman’s dual-system thinking model~\cite{kahneman2002maps} (see Table~\ref{tab: dual system thinking}) to illustrate that the same modalities can yield varying or even contradictory results depending on the type of reflective nudge applied. Additional themes are also derived from participants' responses. The notation (/10) indicates the count of participants who shared similar observations.

\subsubsection{Speed}

\paragraph{Direct Reflective Nudge (Persona).} Participants valued the \textbf{text} modality for its conciseness and efficiency in quickly conveying different perspectives (8/10). ``\textit{Text is quick} (P05), \textit{fastest to read through} (P02, P05, P07) and that the \textit{speed of acquiring information from text is the fastest} (P01).'' Similarly, participants found \textbf{image} effective for quickly conveying perspectives and emotions (5/10):  ``\textit{The images included all the necessary elements to clearly make the point, allowing me to understand the perspective immediately} (P09).'' Notably, for participants whose \textbf{first language is not English}, images were particularly beneficial. ``\textit{Image triggers my reflection faster due to my language barrier. Since English is not my first language, I can quickly understand the content and perspectives from an image using my own intuition, whereas text requires more effort to read and comprehend} (P03).'' Figure~\ref{fig: speed} shows a summary of participant preferences for each modality in terms of speed.

\paragraph{Indirect Reflective Nudge (Storytelling).} 
Feedback on the speed of conveying perspectives for indirect reflective nudge differs notably from direct reflective nudge: \textbf{video} was deemed as the most time-efficient and effective as it encompasses text, image and audio (6/10): ``\textit{Video is the most direct for me, as it allows me to quickly understand and process the content in a shorter time period} (P13).''
By contrast, participants had mixed opinions on the efficiency of the \textbf{text} modality for conveying perspectives. Some found text to be quick and efficient (5/10), while others did not share this view (5/10). Those who appreciated text cited their ability to quickly read and process textual information: ``\textit{It is much faster for me to read, so I spend lesser time understanding the contents. In that sense, text is more efficient and helps me to reach the stage of self-reflection faster.}'' While others expressed concerns about its time-consuming nature. P16 said, ``\textit{I wouldn't want to read long chunks of text}.'' % and P20 commented, ``\textit{Reading long chunks of text slows me down and is too time-consuming.}'' P19 added, ``\textit{I need to go through each sentence several times to fully understand the text, making it very time-consuming.}'' Overall, participants who found text inefficient noted that they would engage with it only if they had ample time (P11, P13), highlighting the reliance of text on the user's time availability.
Two participants found \textbf{images} time-consuming because they were positioned above the text, causing them to scan the images before reading the text. ``\textit{I felt like I kept going back and forth — I was reading the text and trying to match it with the image, which take longer for me to understand the content as a whole} (P19).''

Feedback on the \textbf{audio} modality was consistent with direct reflective nudge. No participants described audio as efficient; in fact, some found it slow (4/10): ``\textit{I find the pace of the audio quite slow; the narration drags on} (P12).''
%while another commented, ``\textit{Audio relies on the speed of the narration, so the information is transmitted slowly}'' (P14). Figure~\ref{fig: speed} shows a summary of participant preferences for each modality in terms of speed.

\subsubsection{Depth of Self-Reflection}

\begin{figure*}[!htbp]
  \centering
  \includegraphics[width=.8\textwidth]{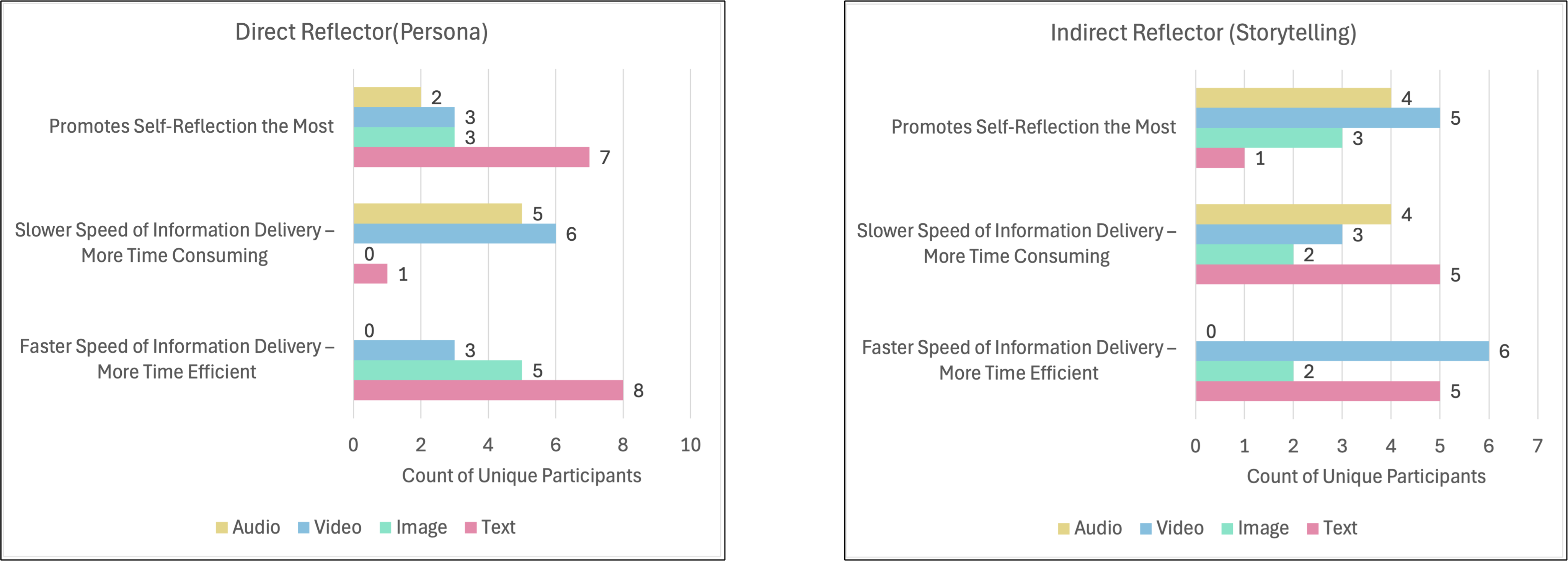}
  \caption{Count of participants who shared the same feedback on the depth of self-reflection and speed of delivery for the four modalities of each reflective nudges: direct (persona) - left and indirect (storytelling) - right.}
  \label{fig: speed}
  \Description{Bar graph showing the count of participants who shared the same feedback on the depth of self-reflection and the speed of delivery for the four modalities of each reflective nudges: direct (persona) on the left and indirect (storytelling) on the right. The bar graph shows that text is reflected to be the most time efficient and enhances self-reflection the most for persona. Whereas, video is reflected to be the most time efficient and enhances self-reflection for storytelling.}
\end{figure*}

\paragraph{Direct Reflective Nudge (Persona)}
Majority of the participants (7/10) found that \textbf{text} facilitated self-reflection more effectively than other modalities for the direct reflective nudge. ``\textit{Text is the most effective medium for reflection because it helps me reconstruct and clarify my thoughts} (P01).'' 
%As one participant noted, ``\textit{Text aids my self-reflection because I retain more information while reading, which helps me remember and think deeply about the topic}” (P05). Another participant mentioned that text offered more details and insights, which supported his thinking process and enhanced self-reflection (P06).
For \textbf{video}, some participants found that while they were not the most efficient in conveying information for direct reflective nudges, they provided more time for self-reflection (3/10). As P01 noted, ``\textit{Although the videos [...] conveyed perspectives slowly, they allowed more time for self-reflection.}'' % P08 added, ``\textit{The video condenses information into short-form content, giving me more time to reflect}''. Similarly, P10 remarked, ``\textit{The slower pacing of the videos compared to text helped me to be more mindful in my comments and facilitated deeper self-reflection}''.
For \textbf{image}, some participants found it beneficial for self-reflection by simplifying complex ideas (3/10). As P10 explained, ``\textit{Images enhance my self-reflection by presenting symbolic meanings that prompt deeper personal interpretation. A powerful image can quickly establish a connection, making it easier for me to engage with and reflect on the topic.}''
%Additionally, images can distill complex ideas into more manageable elements, allowing me to focus on key aspects and think more deeply about the subject}''.
For \textbf{audio}, two participants felt it enhanced self-reflection by simulating a conversational experience, as if interacting with a real person. % ``\textit{Audio facilitates self-reflection because I can hear the speaker’s emotions and tone, helping me to connect with the speaker and making me to think differently about the topic. This would also change how I engage online}'' (P09). 
P02 noted, ``\textit{Audio adds meaning to the text by conveying emotions and expressions that I can’t grasp from reading text alone, helping me to better understand the speaker’s intent.}'' Figure~\ref{fig: speed} shows a summary of participant preferences for each modality in terms of depth of self-reflection.

\paragraph{Indirect Reflective Nudge (Storytelling).} Half of the participants (5/10) found that \textbf{video} enhances self-reflection by making the content more relatable and connecting with their personal experiences: ``\textit{The video portrays daily life well, allowing me to compare it with my own experiences, which prompted deeper reflection} (P20).'' Additionally, participants found that videos aid in clarifying their stance and feelings on issues.
%P14 further added, ``\textit{Videos help me self-reflect by making it easier to empathize with the characters, deepening my reflection}''.: ``\textit{Videos are helpful in figuring out my feelings or stance on an issue when I'm unsure}'' (P12).
In contrast to direct reflective nudge, where text was highly valued for its role in self-reflection, only one participant found \textbf{text} effectively facilitated self-reflection for indirect reflective nudge. %``\textit{The greater autonomy in interacting with the text contributed to enhancing my self-reflection}'' (P17).
For \textbf{image}, a few participants (3/10) noted that the memorable nature of images enhances self-reflection by aiding in content recall. As P19 put it, ``\textit{What helps me self-reflect more is the modality’s ability to make the content memorable. Images stand out for me because I can vividly remember the content, which aids in deeper self-reflection}.''

Compared to direct reflective nudge, a higher proportion of participants found \textbf{audio} effective for self-reflection (4/10), citing similar reasons. Specifically, they noted that the presence of a voice conveyed tone and emotion, aiding their reflection: ``\textit{Hearing the voice helps me understand the character’s emotions and plight better which enhances my self-reflection} (P13).'' 
%Audio also facilitated their thought process: ``\textit{Listening to the audio allowed me to quickly agree or disagree and refine my thoughts as the audio continues to play}'' (P15). 
Additionally, audio enabled participants to immerse themselves in the role of the main character: ``\textit{Audio lets me close my eyes and imagine living as each person, which stimulates my self-reflection} (P11).'' Figure~\ref{fig: speed} shows a summary of participant preferences for each modality in terms of the depth of self-reflection.

\subsubsection{Processing (Both Nudges)}
Regardless of the type of reflective nudge, participants from both groups found that \textbf{videos} and \textbf{images} helped them grasp the \textbf{broader context} or overall picture, whereas \textbf{text} allowed for a deeper understanding of \textbf{finer details}. Videos and images provided a higher-level overview and abstraction, making it easier to understand the main idea (P06, P09). Conversely, \textbf{text} was appreciated for its ability to convey detailed concrete information and insights, offering a more comprehensive understanding of the topic (P10, P06).
%For instance, one participant noted, ``\textit{From the image alone, I could grasp the main gist but missed the finer details}'' (P19; Indirect). 

\subsubsection{Effort (Both Nudges)}
Participants from both groups found that \textbf{audio} demanded significant \textbf{concentration, focus and mental effort}, summarized by the following statement: ``\textit{It’s very hard for me to engage with audio alone} (P04).'' %``\textit{Listening while tracking the contents of the audio is challenging}'' (P06; Direct), and ``\textit{I struggled with audio most due to limited visual information}'' (P03; Direct), reflected this difficulty. 
Participants also mentioned needing to stay actively focused on the audio to avoid missing content (P05) and finding it hard to absorb information (P11).
%, thus requiring substantial concentration and effort (P13, P20; Indirect).
\textbf{Video}, on the other hand, was described as facilitating content processing: ``\textit{Video makes it easier for me to process the content} (P13).''

\subsubsection{Nature (Both Nudges)}
Participants noted that \textbf{text} and \textbf{video} are more familiar and accessible due to \textbf{daily exposure}. Text is described as traditional and conventional, making it easier to absorb information (P08, P12). It is considered the most accessible modality (P04) and is frequently used, leading to greater familiarity (P10). Similarly, videos were likened to short-form content found on platforms like TikTok and YouTube Shorts, making them easier to digest as they are familiar with it (P04, P08, P14).
%Participants thus appreciated this similarity to social media content (P19, P18; Indirect).
In contrast, \textbf{audio} was less favored due to its \textbf{limited natural exposure}. Those who did prefer it attributed this preference to their inherent tendency towards auditory learning. Participants in both groups generally expressed disinterest or discomfort with audio, preferring visual stimuli instead. 
%Feedback such as, ``\textit{I am generally not an auditory person}'' (P02; Direct, P19; Indirect) and ``\textit{I tend to doze off with audio, so audio is not for me}'' (P02; Direct) reflected this sentiment. Contrarily, some participants who identified as auditory learners found it easier to absorb information through listening (P11; Indirect).

\subsubsection{Information Retention (Both Nudges)}
Both \textbf{video} and \textbf{audio} exhibit slightly \textbf{lower information retention} compared to the more static modalities of text and images. Participants noted that ``\textit{for both video and audio, I don’t retain as much information compared to text} (P05)'', and that ``\textit{retention rates for recalling specific details from video and audio are lower, as I tend to forget earlier parts while processing new information} (P18).''  Another participant mentioned, ``\textit{videos convey a lot quickly, but I forget the details just as quickly because the information is presented in a short time} (P19).'' In contrast, \textbf{text} and \textbf{images} are described as \textbf{more memorable} (P01, P05, P18, P19).
%, with participants noting, ``\textit{I can recall the contents better because they stick in my head}'' (P20).

\subsubsection{Engagement (Both Nudges)}
\textbf{Text} was generally reported as the \textbf{least engaging}, while \textbf{video} was viewed as \textbf{highly engaging} by most participants in both groups. Videos and \textbf{images} were praised for their \textbf{attention-grabbing} qualities (P01, P13, P14, P20). Videos were particularly noted for being interactive, fun and interesting (P01, P14, P15).
%and were found less boring than plain audio, static images, or text (P01, P02, P09; Direct) The dynamic nature of video scenes also kept viewers engaged (P20; Indirect). 
In contrast, text was described as dry and boring (P01, P15, P18, P20). Participants found it less captivating, especially in an era dominated by multimedia content (P10). %and noted that there is nothing \textit{'special'} about plain text (P09; Direct).

\subsubsection{Users' Autonomy (Both Nudges)}
\textbf{Video} and \textbf{audio} generally offered the \textbf{lowest level of user autonomy}: users noted that they must watch or listen to the entire content without control over the pace (P03, P05). % ``\textit{Videos and audios have fixed playing time so I have less control on how much time I want to spend, but for image and text, they just depends on how fast I can go}'' (P12, P14, P18; Indirect). 
In contrast, \textbf{text} was reported to provide the \textbf{highest level of autonomy}. Participants appreciated that they could adjust their reading speed and review information at their own pace (P01, P03, P16, P17).

\subsubsection{Credibility and Trust in AI-Generated Content (Both Nudges)}
Participants generally reported \textbf{lower credibility} for \textbf{images} compared to audio, video, and text. Some expressed skepticism toward AI-generated images (P04, P14), noting that they appeared unrealistic and less human-like (P01, P08). In contrast, video and audio were seen as more credible due to their human-like elements, such as real human presence in videos and realistic voices in audio (P01).

\subsubsection{Usage Scenarios (Both Nudges)}
\label{sec: usage scenarios}
In general, participants found that \textbf{video} is beneficial for those with little to \textbf{no prior knowledge} of the topic. It helps provide context and explains the issue from multiple angles, making it useful for learning and reflection (P04, P06, P11). For \textbf{images}, they are effective in \textbf{triggering self-reflection for those with some prior knowledge on the topic}. Images are attention-grabbing and can engage users by capturing their focus (P06, P13, P20). 
%``\textit{For an effective reflection, I need something that is more attractive, so I am more likely to engage with images as they capture my attention}'' (P14; Indirect). Images thus serve as a good starting point for self-reflection, helping to kick-start the process (P08, P09; Direct, P11; Indirect). 
\textbf{Text} is primarily used by individuals who already have a \textbf{deep interest or prior knowledge of the topic} (P05, P06). For these users, text provides detailed information that supports reflection without needing additional attention-grabbing elements (P06). Lastly, an overwhelming number of participants mentioned that \textbf{audio} is suited for \textbf{multitasking} and \textbf{on-the-go reflection}, as it allows them to reflect while engaging in other activities.

\subsection{Summary of the Results}
Our results revealed how different modalities of reflective nudges interact with the dual-system thinking framework. Specifically, we observed a clear preference trend: text was favored for direct reflective nudges due to its speed and autonomy, enhancing self-reflection, albeit being less engaging. In contrast, video was preferred for indirect reflective nudges because of its speed, engagement and ability to convey the main idea effectively, despite its lower autonomy. Appendix Figures~\ref{fig: butterfly chart text} - ~\ref{fig: butterfly chart audio} present a butterfly chart summarizing the qualitative feedback for each modality across both nudge types. 
\section{Study 2: Assessing the Impacts of Multimodal Reflection Nudges on Deliberativeness}
In study 1, we identified the preferred modalities for each type of reflective nudge. To further understand how these modalities influence deliberativeness, we conducted study 2 to assess their impact on deliberative quality.

\subsection{Independent Variables and Experimental Design}
We used the same 2 $\times$ 4 design with Reflective Nudge: \{Direct (Persona), Indirect (Storytelling)\} and Modality: \{Text, Image, Video, Audio\}. Both independent variables were between-subject. Thus, participants were randomly assigned to one of the eight experimental conditions. 

The experimental interface retained the same design used in study 1 (Figure~\ref{fig: features}), but for this study, the \textit{Reflect} feature was limited to a single modality. 

Similar to study 1, we do not compare direct and indirect reflective nudges; instead, we evaluate the effects of different modalities within each nudge separately.

\subsection{Dependent Variables}
As deliberativeness is multi-dimensional, we operationalized it through five measurements as done by previous work~\cite{yeo2024help}: argument repertoire, argument diversity, rationality (opinion expression), rationality (justification level) and constructiveness as discussed in section~\ref{sec: measurements}.

All five metrics were derived from a content analysis of participants' responses by two coders. Both coders were PhD students with respectively 1 and 4 years of experience using content analysis. Cohen's Kappa was used to determine the agreement between the two coders’ judgments, with individual scores reported below. Kappa scores for all metrics were above the satisfactory threshold of 0.70~\cite{viera2005understanding, mchugh2012interrater}. 

The dependent variables are coded as follows:
\begin{itemize}
    \item \textit{Argument repertoire} ($\kappa = 0.915$) is the number of non-redundant arguments regarding each position of the discussion topic. The ideas produced along the two positions were combined.
    \item \textit{Argument diversity} ($\kappa = 0.915$) was coded by counting the number of unique themes present in the entire response. A higher diversity count indicates more varied perspectives present in the participant's responses~\cite{anderson2016all, gao2023coaicoder}.   
    \item \textit{Rationality (opinion)} ($\kappa = 0.872$) captures whether opinions are expressed or information is provided in the response. This was coded with two levels: 1) no opinions were expressed, rather, information was provided (score of 0); 2) an opinion, personal assertion or a claim was made (score of 1). This also includes evaluation, a personal judgment or assertion. 
    \item \textit{Rationality (justification level)} ($\kappa = 0.867$) captures the degree to which reasons are used to justify one's claims. This were coded at four levels: 1) no justification was provided (score of 0); 2) an inferior justification was made - this indicates that the opinion is supported with a reason in an associational way such as through personal experiences or an incomplete inference was given (score of 1); 3) a qualified justification was made when there is a single complete inference provided in the opinion (score of 2); 4) a sophisticated justification was made when at least two complete inferences was provided (score of 3).  
    \item \textit{Constructiveness} ($\kappa = 0.830$) captures the degree of balance within an opinion. This was coded at two levels: 1) the opinion is one-sided (score of 0); 2) the opinion is two-sided when multiple perspectives and viewpoints are presented (score of 1). 
\end{itemize}

\subsection{Power Analysis}
We conducted a power calculation for a eight-group ANOVA study seeking a medium effect size (0.30) according to Cohen’s conventions, at 0.80 observed power with an alpha of 0.05, giving $N=25$ per experimental condition, hence we recruited 200 participants. 

\subsection{Participants and Ethics}
A total of 200 participants were recruited through Amazon Mechanical Turk. Refer to Appendix Table~\ref{tab: st2-demo} on the breakdown of the demographic profile in each experimental condition. We ensured that the demographic profiles across the eight conditions were similar so as to control for any fixed effects resulting from the differences in demographic factors. Similar to study 1, we got ethics approval from our local IRB and reimbursed participants at an appropriate rate.

\subsection{Procedure and Task}
The procedure for this study closely follows that of study 1 outlined in section~\ref{sec: procedure}. In the pre-task phase, we gathered data on demographics and administered four questionnaires corresponding to each of the four covariates, as detailed in section~\ref{sec: Covariates}. Instructions on reflection were then provided to the participants.

In the main task, participants were instructed to utilize the \textit{Reflect} feature, which exclusively presents prompts from a single modality. The topic remained the same as study 1. We maintain consistency in the topic across the two studies to draw robust comparisons between different modalities and guarantee the internal validity of our results~\cite{cahit2015internal}.

In the post-task phase, participants completed a short survey in which they could provide feedback on the modality they engaged with.
\section{Results (Study 2)}
 
\subsection{Quantitative Results}
\label{sec: quantitative}
Before analyzing our data, we plotted a box plot (box and whisker plot) to visually show the dispersion of our data and to identify any potential outliers~\cite{schwertman2004simple, dawson2011significant}. No abnormalities in the data were observed. 

ANCOVA was then used to identify main effects while controlling for the four covariates. We applied pairwise t-tests with Benjamini-Hochberg correction for post-hoc comparisons for the eight measures of deliberativeness. We did not use Bonferroni correction, as its conservative approach leads to high rates of false negatives when done with large number of comparisons~\cite{thissen2002quick}. Instead, we relied on Benjamini-Hochberg as it minimizes the problem~\cite{nakagawa2004farewell} while still accounting for multiple comparisons. We report effects of covariates only where they are significant.

Results are summarized in Tables~\ref{tab: summary quantitative direct} and ~\ref{tab: summary quantitative indirect}.

\subsubsection{Argument Repertoire}

\paragraph{Direct Reflective Nudge} We found a significant main effect of \emph{Modality} on \emph{Argument Repertoire} for direct reflective nudge ($F_{3,86} = 4.87$, $p<.01$). There was a statistical difference between \emph{Video} ($M = 3.60$ arguments) and \emph{Text} ($M = 2.32$ arguments), \emph{Image} ($M = 2.56$ arguments) and \emph{Audio} ($M = 2.60$ arguments) (all $p<.05$). %Results are summarized in Figure~\ref{fig: argument repertoire}.

\paragraph{Indirect Reflective Nudge} No significant main effect of \emph{Modality} on \emph{Argument Repertoire} was found for indirect reflective nudge ($p=.09$).
%Results are summarized in Figure~\ref{fig: argument repertoire}.

% \begin{figure*}[!htbp]
%   \centering
%   \includegraphics[width=.8\textwidth]{figures/Argument_Repertoire.png}
%   \caption{Argument Repertoire for direct reflective nudge (left) and indirect reflective nudge (right). We report the results of the ANCOVA test, and pairwise comparisons with BH correction if any, where * : p < .05, ** : p < .01.}
%   \label{fig: argument repertoire}
%   \Description{Box plot describing argument repertoire for the four modalities for direct reflective nudge (left) and indirect reflective nudge (right). The X-axis shows the different modalities and the Y-axis shows the number of arguments.}
% \end{figure*}

\subsubsection{Argument Diversity}

\paragraph{Direct Reflective Nudge} No significant main effect of \emph{Modality} on \emph{Argument Diversity} was found for direct reflective nudge ($p=0.085$). However, individual's inherent reflecting styles, particularly a lower preference for text, was associated with higher argument diversity ($p<.05$). %Results are summarized in Figure~\ref{fig: argument diversity}.

\paragraph{Indirect Reflective Nudge} We found a significant main effect of \emph{Modality} on \emph{Argument Diversity} for indirect reflective nudge ($F_{3,86} = 3.41$, $p<.05$). We observed pairwise differences between \emph{Video} ($M = 5.52$ themes) and \emph{Image} ($M = 3.72$ themes, $p<.05$) only.
%Results are summarized in Figure~\ref{fig: argument diversity}.

% \begin{figure*}[!htbp]
%   \centering
%   \includegraphics[width=.8\textwidth]{figures/Argument_Diversity.png}
%   \caption{Argument Diversity for direct reflective nudge (left) and indirect reflective nudge (right). We report the results of the ANCOVA test, and pairwise comparisons with BH correction if any, where * : p < .05, ** : p < .01.}
%   \label{fig: argument diversity}
%   \Description{Box plot describing argument diversity for the four modalities for direct reflective nudge (left) and indirect reflective nudge (right). The X-axis shows the different modalities and the Y-axis shows the number of argument diversity.}
% \end{figure*}

\subsubsection{Rationality (Opinion)}

\paragraph{Direct Reflective Nudge} No significant main effect of \emph{Modality} on \emph{Opinion} was found ($p=0.224$). However, a higher tendency to self-reflect ($p<.01$), lower inherent preference for audio ($p<.01$) and reduced external exposure to images ($p<.05$) were associated with higher expression of opinions. %Results are summarized in Figure~\ref{fig: opinion}.

\paragraph{Indirect Reflective Nudge} We found a significant main effect of \emph{Modality} on \emph{Opinion} for indirect reflective nudge ($F_{3,86} = 4.61$, $p<.01$). Overall, we found differences between \emph{Text} ($M = 0.48$) versus \emph{Image} ($M = 0.76$), \emph{Video} ($M = 0.84$) and \emph{Audio} ($M = 0.84$) (all $p<.05$). %Results are summarized in Figure~\ref{fig: opinion}.

% \begin{figure*}[!htbp]
%   \centering
%   \includegraphics[width=.8\textwidth]{figures/Opinion.png}
%   \caption{Rationality (Opinion) for direct reflective nudge (left) and indirect reflective nudge (right). We report the results of the ANCOVA test, and pairwise comparisons with BH correction if any, where * : p < .05, ** : p < .01.}
%   \label{fig: opinion}
%   \Description{Box plot describing opinion expression for the four modalities for direct reflective nudge (left) and indirect reflective nudge (right). The X-axis shows the different modalities and the Y-axis shows the level of opinion expression.}
% \end{figure*}

\subsubsection{Rationality (Justification Level)}

\paragraph{Direct Reflective Nudge} No significant main effect of \emph{Modality} on \emph{Justification Level} was observed ($p = 0.571$). However, higher inherent aural preferences and lower inherent text preferences as well as greater external exposure to videos (all $p<.05$) are associated with higher levels of justification. % Results are summarized in Figure~\ref{fig: justification level}.

\paragraph{Indirect Reflective Nudge} We found a significant main effect of \emph{Modality} on \emph{Justification Level} for indirect reflective nudge ($F_{3,86} = 3.39$, $p<.05$). There was a statistical difference between \emph{Video} ($M = 2.32$) and \emph{Image} ($M = 1.44$, $p<.05$).
% Results are summarized in Figure~\ref{fig: justification level}.

% \begin{figure*}[!htbp]
%   \centering
%   \includegraphics[width=.8\textwidth]{figures/Justification_Level.png}
%   \caption{Rationality (Justification Level) for direct reflective nudge (left) and indirect reflective nudge (right). We report the results of the ANCOVA test, and pairwise comparisons with BH correction if any, where * : p < .05, ** : p < .01.}
%   \label{fig: justification level}
%   \Description{Box plot describing justification level for the four modalities for direct reflective nudge (left) and indirect reflective nudge (right). The X-axis shows the different modalities and the Y-axis shows the level of justification.}
% \end{figure*}

\subsubsection{Constructiveness}

\paragraph{Direct Reflective Nudge} No significant main effect of \emph{Modality} on \emph{Constructiveness} was observed ($p = 0.126$). However, a higher inherent \emph{aural} ($p<.01$) and \emph{image} ($p<.05$) preferences as well as reduced external exposure to text ($p<.05$) are associated with higher constructiveness.
%Results are summarized in Figure~\ref{fig: constructiveness}.

\paragraph{Indirect Reflective Nudge} No significant main effect of \emph{Modality} on \emph{Constructiveness} was observed ($p = 0.169$). 
%Results are summarized in Figure~\ref{fig: constructiveness}.

% \begin{figure*}[!htbp]
%   \centering
%   \includegraphics[width=.8\textwidth]{figures/Constructiveness.png}
%   \caption{Rationality (Justification Type) for direct reflective nudge (left) and indirect reflective nudge (right). Error bars show .95 confidence intervals. We report the results of the ANCOVA test, and pairwise comparisons with BH correction if any, where * : p < .05, ** : p < .01.}
%   \label{fig: constructiveness}
%   \Description{}
% \end{figure*}

\begin{table*}[!htbp]
\caption{Deliberative quality across the four modalities for direct reflective nudge (Persona). $\alpha$,$\beta$,$\gamma$ show significant pairwise differences. n.s.: not significant, *: $p<.05$, **: $p<.01$, etc...}
\label{tab: summary quantitative direct}
\begin{tabular}{cccccc}
\toprule
\multirow{3}{*}{\textbf{Deliberativeness}} & \multirow{3}{*}{\textbf{$p$}} & \multicolumn{4}{c}{\textbf{Modality}} \\ \cline{3-6}
 & & \textbf{\begin{tabular}[c]{@{}c@{}}Text\\      ($M \pm S.D.$)\end{tabular}} & \textbf{\begin{tabular}[c]{@{}c@{}}Image\\      ($M \pm S.D.$)\end{tabular}} & \textbf{\begin{tabular}[c]{@{}c@{}}Video\\      ($M \pm S.D.$)\end{tabular}} & \textbf{\begin{tabular}[c]{@{}c@{}}Audio\\      ($M \pm S.D.$)\end{tabular}} \\
\midrule
Argument Repertoire & ** & 2.32 ± 1.07$^\alpha$ & 2.56 ± 1.00$^\beta$ & 3.60 ± 1.94$^\alpha$$^\beta$$^\gamma$ & 2.60 ± 1.41$^\gamma$ \\
Argument Diversity & n.s. & 4.08 ± 1.68 & 5.08 ± 1.75 & 5.48 ± 2.63 & 4.64 ± 1.80 \\
Rationality (Opinion Expression) & n.s. & 0.64 ± 0.49 & 0.84 ± 0.37 & 0.60 ± 0.50 & 0.72 ± 0.46 \\
Rationality (Justification Level) & n.s. & 2.20 ± 0.91 & 2.08 ± 0.81 & 2.20 ± 0.82 & 1.92 ± 0.86 \\
Constructiveness & n.s.& 0.28 ± 0.46 & 0.20 ± 0.41 & 0.48 ± 0.51 & 0.36 ± 0.49 \\
\bottomrule
\end{tabular}
\Description{A table summarizing the deliberative quality for direct reflective nudge across the four modalities and across the dependent variables. The table is a summary of the other Figures in the section.}
\end{table*}

\begin{table*}[!htbp]
\caption{Deliberative quality across the four modalities for indirect reflective nudge (Storytelling). $\alpha$,$\beta$,$\gamma$ show significant pairwise differences. n.s.: not significant, *: $p<.05$, **: $p<.01$, etc...}
\label{tab: summary quantitative indirect}
\begin{tabular}{cccccc}
\toprule
\multirow{3}{*}{\textbf{Deliberativeness}} & \multirow{3}{*}{\textbf{$p$}} & \multicolumn{4}{c}{\textbf{Modality}} \\ \cline{3-6}
  & & \textbf{\begin{tabular}[c]{@{}c@{}}Text\\      ($M \pm S.D.$)\end{tabular}} & \textbf{\begin{tabular}[c]{@{}c@{}}Image\\      ($M \pm S.D.$)\end{tabular}} & \textbf{\begin{tabular}[c]{@{}c@{}}Video\\      ($M \pm S.D.$)\end{tabular}} & \textbf{\begin{tabular}[c]{@{}c@{}}Audio\\      ($M \pm S.D.$)\end{tabular}} \\
\midrule
Argument Repertoire & n.s. & 2.76 ± 1.13 & 2.60 ± 1.26 & 2.88 ± 1.96 & 2.84 ± 1.28 \\
Argument Diversity & * & 4.24 ± 2.07 & 3.72 ± 2.15$^\alpha$ & 5.52 ± 2.40$^\alpha$ & 4.92 ± 1.82 \\
Rationality (Opinion Expression) & ** & 0.48 ± 0.51$^\alpha$$^\beta$$^\gamma$ & 0.76 ± 0.44$^\alpha$ & 0.84 ± 0.37$^\beta$ & 0.84 ± 0.37$^\gamma$ \\
Rationality (Justification Level) & * & 1.84 ± 1.11 & 1.44 ± 1.08$^\alpha$ & 2.32 ± 0.85$^\alpha$ & 2.08 ± 1.08 \\
Constructiveness & n.s. & 0.16 ± 0.37 & 0.24 ± 0.44 & 0.32 ± 0.48 & 0.44 ± 0.51 \\
\bottomrule
\end{tabular}
\Description{A table summarizing the deliberative quality for indirect reflective nudge across the four modalities and across the dependent variables. The table is a summary of the other Figures in the section.}
\end{table*}

\subsubsection{Summary of Results}
Overall deliberative quality across the five measurements is summarized in Tables~\ref{tab: summary quantitative direct} and \ref{tab: summary quantitative indirect} for direct and indirect reflective nudges, respectively. A detailed analysis reveals that \emph{Video} generally emerges as the most effective modality for the direct reflective nudge (persona), particularly excelling in Argument Repertoire. For the indirect reflective nudge (storytelling), similarly, \emph{Video} leads in Opinion Expression, Argument Diversity, and Justification Level, often outperforming \emph{Image} significantly. This suggests that video enhances user engagement and expression, thereby improving deliberativeness. In contrast, \emph{Text} is least effective for Opinion Expression in the context of indirect reflective nudges (storytelling), suggesting that lengthy text may not engage users as effectively in articulating their opinions.

\subsection{Subjective Feedback}
\label{sec: qualitative}
We present the findings from the thematic analysis of the post-task feedback, where numbers in parentheses indicate the frequency of recurring feedback across both nudge types. Notably, no specific dislikes were reported for text or images in either reflective nudge type.

\subsubsection{General Strengths of Reflect}
Participants demonstrated strong appreciation for the system across all modalities in both reflective nudges. They appreciated its thought-provoking nature, which facilitated critical thinking and led to a deeper understanding of their own feelings and thoughts on the issue (40). It helped clarify and refine their perspectives (10), encourage deep self-evaluation, foster self-awareness (21), and prompt reflection on personal values and the impact of their opinions (8). 
%Additionally, by connecting to their own experiences, the system resonated with participants, making the content more relatable (15), while also reinforcing their understanding of the topic (21). Participants also mentioned that they felt emotionally supported as they were able to explore complex emotions and moral dilemmas (14). 
With its ability to present multiple viewpoints and perspectives that might have otherwise been overlooked (29), participants found the system both interesting and useful (34). 
%Its ease of use, clear and intuitive design, and simplicity made it highly accessible (8). 

\subsubsection{Strengths of each Modality in Direct Reflective Nudge}
Participant appreciated that \textbf{text} was straightforward and concise (P12). For \textbf{images}, participants noted that they helped break down complex information, making it easier to grasp, retain and connect with the content (P26).

Participants found \textbf{videos} particularly engaging (P51, P57) due to their multi-sensory approach of combining visual and auditory elements (P51, P68, P70, P74). This blend elicited deeper emotional and cognitive responses (P54, P55, P57): ``\textit{Videos fostered both emotional and intellectual engagement by leveraging storytelling, visuals, and music, making it more memorable and impactful} (P68).''
%Additionally, videos offered vivid, real-world examples that made abstract concepts more tangible and relatable (P74).

For \textbf{audio}, participants appreciated that the voice sets a calm, reflective atmosphere for them to engage deeply with their thoughts and emotions (P77), enhancing the reflective experience (P78, P84).
%, with the voice evoking emotions and memories, facilitating a deeper connection to the reflection process (P82, P83). 
%The immersive nature of audio also helps participants focus inwardly without distractions, allowing them to fully engage with their thoughts and feelings (P85, P89, P91, P95). 
Additionally, some participants preferred audio over reading, noting that it kept them more engaged: ``\textit{I liked that I did not have to read anything. Listening to the audio gave me more motivation to explore multiple opinions, whereas with text, I would have only read one or two [chunks] at most} (P94).''

\subsubsection{Strengths of each Modality in Indirect Reflective Nudge}
Participants appreciated the \textbf{text} modality for its conciseness and ease of engagement (P111). For \textbf{images}, participants noted the high quality of the AI-generated visuals, describing them as ``well done'' (P126), and found that they stimulated their imagination (P144). For \textbf{both text and images}, participants found the well-developed stories particularly memorable (P124), with images resonating deeply and prompting significant reflection (P134). 
%In both modalities, the relatable journeys and dilemmas faced by the characters encouraged participants to reflect on their own values and decisions, as well as how these experiences influenced their personal growth (P114, P116, P117, P118, P142).

For \textbf{videos}, participants provided similar feedback as with the direct reflective nudge. They found videos particularly engaging (P152) due to the multi-sensory combination of visual and auditory elements, which helped them better articulate their thoughts and feelings (P152). 
Similarly, for \textbf{audio}, participants shared feedback consistent with the direct reflective nudge. Fourteen participants reported they liked the calm voice which fostered more profound reflection by reducing distractions.
%This blend also allowed participants to process complex ideas more effectively (P152, P169). The videos also made the scenarios more relatable and vivid, offering real-world examples (P154, P160, P170, P171, P172, P174), which helped participants connect emotionally and intellectually with the thought-provoking content (P154, P159, P167, P170, P173, P174).
%For  Again, (P176, P182, P197, P198) and enhancing emotional connection and engagement with the content (P176, P178). Additionally, P192 noted that ``\textit{the voice added a dynamic and reflective quality to the topic, infusing it with personality and depth, enriching my contemplation of the subject}''.

\subsubsection{Critiques of each Modality in Direct Reflective Nudge}
For \textbf{video}, one participant expressed a dislike for computer-generated narration (P52). Additionally, two participants mentioned that the lack of interactivity in the videos led to passive consumption of information rather than active engagement which may limit the depth of reflection (P70, P73).

For \textbf{audio}, four participants expressed that it was not for them: ``\textit{Audio can certainly aid in reflection, but it might not always hit the mark for every individual} (P84).'' % and ``\textit{It may not always suit everyone’s reflective needs or styles}'' (P76). 
%This underscores the importance of finding a reflective practice that aligns with individual reflective preferences and needs (P80, P92). 
Additionally, three participants found the audio voices too robotic, noting a desire for a more human-like tone to enhance credibility (P79, P83, P94).

\subsubsection{Critiques of each Modality in Indirect Reflective Nudge}
%For \textbf{video}, one participant felt that some videos felt overly condensed, making it difficult to fully absorb the information and engage in meaningful reflection (P152). 

As for \textbf{audio}, two participants found that it became less engaging over time, suggesting that subtle changes or more variety were needed for continued engagement: ``\textit{After a while, it became less engaging, and I found my mind drifting away from the reflective process. A bit more variety or subtle shifts in the audio could have kept my focus and enhanced the overall reflective experience} (P180).''
\section{Discussion}
In this section, we address \textbf{RQ2: How does the modality of a reflective nudge affect the quality of deliberation?} by synthesizing quantitative data from our objective measures (see section~\ref{sec: quantitative}), qualitative feedback from participants (see section~\ref{sec: qualitative}) and insights from study 1 to provide a comprehensive view of how different modalities within different reflective nudges impact deliberativeness. We also examine how our findings corroborate and enrich previous research, discussing the broader implications of enhancing deliberativeness on online deliberation platforms.

\subsection{Tailoring Modalities to Suit Different Types of Reflective Nudges}
Our results from study 1 demonstrate that text is most preferred for direct reflective nudges, while video is favored for indirect reflective nudges. Study 2's quantitative analysis confirms that video significantly enhances deliberativeness for indirect reflective nudges, increasing argument diversity, opinion expression, and justification levels. Conversely, text performs the worst for opinion expression, highlighting that the same modality on different nudge types significantly impacts deliberative quality.

This goes to show that text-based reflective nudges are most effective when they are short and straightforward, aligning with earlier work that lengthy textual information can cause cognitive overload and reduce engagement~\cite{sweller1988cognitive}. For complex reflective tasks, video offers a better alternative, suggesting that \textbf{indirect reflective nudges benefit more from direct modalities} like video, which maintain engagement and convey ideas effectively.

The preference for video in indirect reflective nudges can be attributed to its capacity of multi-sensory engagement, allowing it to deliver complex concepts more directly and engagingly~\cite{clark2023learning} as highlighted by the qualitative feedback in both studies (sections~\ref{sec: qualitative for study 1} and \ref{sec: qualitative}), supporting Mayer’s findings~\cite{mayer2005cambridge} that video outperforms images and text in learning contexts. Additionally, Dale’s~\cite{dale1969audiovisual} cone of learning highlights the inherent concreteness of videos compared to more abstract modalities like pictures. This concreteness likely contributes to their effectiveness in promoting reflection by providing richer, more relatable stimuli compared to images or text. Our study extends these findings to reflection in online deliberation, showing that video enhances deliberativeness more effectively than text alone, echoing the multimedia principle~\cite{fletcher2005multimedia}.

Overall, our findings suggest that as the complexity of reflective nudge increases, different modalities may be needed to enhance the reflection process. While text works for concise reflective nudge, richer modalities like video should be considered for more complex nudges. These insights suggest that \textbf{the modality chosen should be carefully aligned with the type of nudge being delivered}, challenging the conventional reliance on text-based approaches and advocating for a multi-modal strategy to better support deliberativeness.

\subsection{Supporting Deliberation with Modalities}

\subsubsection{Importance of Multimodality in Reflection}
Results from studies 1 and 2 revealed that all participants exhibited multimodal preferences, indicating that none had a dominant single preference. This aligns with existing research and theoretical frameworks~\cite{mayer2002multimedia, fleming1992not}, which showed that only a small minority of individuals have single-modal preferences, while the majority are multimodal, spanning bimodal, trimodal, or four-part preferences~\cite{mayer2005cognitive, varklearnVARKResearchWhat}. Notably, the most common is the four-part preference, with 25.4\% of individuals categorized as ``Integrative Multimodal''~\cite{varklearnVARKResearchWhat}. These individuals seek input across all modalities before integrating insights from multiple sources for a more comprehensive grasp of the material. 

Therefore, it is likely that individuals may engage with and benefit from multiple sensory channels for more effective information processing. Studies in educational psychology~\cite{mayer2005cambridge, mayer2002multimedia, fleming1992not, clark2023learning} emphasize the cognitive benefits of multi-sensory engagement, supporting our findings that multimodal approaches can significantly enhance user reflection. By \textbf{leveraging users' multimodal preferences as a catalyst for reflection}, online deliberation platforms can tailor nudges to individual needs more effectively. Integrating multiple modalities into reflective nudges not only caters to diverse user preferences but also leads to more meaningful and impactful deliberation. Ultimately, multimodal approaches ensure that reflective nudges resonate with a broader audience, thereby improving deliberativeness.

\subsubsection{Pairing Modalities - Leveraging the Advantages of Different Modalities for Reflection}
Modalities serve both \textit{interpretative} and \textit{reflective} support, helping users construct opinions while enhancing self-reflection~\cite{reid2003supporting}. Each modality has unique strengths, as evidenced by our qualitative findings from study 1. For instance, participants whose first language was not English, preferred images over text, highlighting the universal accessibility of images and its ability to quickly convey ideas and emotions without language barriers. This underscores the importance of leveraging different modalities not only to enhance inclusivity when language and cultures differs but also to take advantage of their distinct benefits.

Research on temporal contiguity~\cite{moreno1999cognitive} shows that combining narration with animation yields better performance compared to using a single channel, particularly for cognitive tasks~\cite{lee1997effect, paivio2013imagery}. Similarly, Nathan et al.~\cite{nathan1992theory} found pairing verbal and non-verbal materials outperforms narration alone. In our study, participants noted that audio alone, while conversational, was less effective for reflection, lacking sufficient capacity for dual coding~\cite{paivio1975free} and semantic processing (see section~\ref{sec: learning and reflection}). Pairing audio with visuals leverages the \textbf{additive effect}, where combinations like audio with images or text create richer and more effective reflective experiences, amplifying semantic processing and providing a stronger foundation for reflection.

Our findings (see section~\ref{sec: qualitative for study 1}) also illuminate the interaction between modalities and the dual-system thinking (see section~\ref{sec: dual system thinking}). Text and images aligned with System 1, facilitating rapid information delivery, especially for direct reflective nudges. Videos, offering a multi-sensory experience, balanced System 1’s speed with System 2’s reflective depth, particularly for indirect reflective nudges. Parallel processing (System 1) was supported by visual modalities like images and videos, which provided an overarching view of the topic. Serial processing (System 2), tied to detailed and step-by-step analysis, was best facilitated by text, aiding deeper reflective engagement.

By understanding how modalities interact with System 1 and System 2, platforms can design reflective nudges that optimize both intuitive and analytical engagement. Combining modalities strategically can also amplify their strengths, promote inclusivity, and enhance the depth of reflection. Future studies may want to explore combining different modalities to achieve optimal outcomes tailored to specific deliberative environments, enhancing the quality and depth of online discussions.

\subsubsection{Using Modalities to Kick-start Reflection}
Participants highlighted that images effectively initiated their reflection (see section~\ref{sec: usage scenarios}). This aligns with research showing that visual elements, such as images and videos, activate prior knowledge and trigger reflective thinking~\cite{mayer2005cambridge}. By engaging users visually, these modalities provide a foundation that supports more in-depth reflection, allowing users to anchor their initial reflections, which can then be expanded upon when articulating their opinions. Therefore, integrating images and videos early in the reflection process could facilitate more meaningful deliberation.

\subsection{Consider Individual Differences: The Role of Covariates in Multimodal Preferences}
Reflection is a multifaceted process~\cite{mayer2005cambridge} influenced by various factors beyond just the modality used. Our study highlights that covariates significantly shape deliberativeness (see section~\ref{sec: quantitative}). It is essential to account for these individual characteristics when designing reflective nudges, as they determine how users engage with each modality.

Our results are consistent with Mayer et al.~\cite{mayer2002multimedia, mayer1990illustration}, suggesting that modalities should be considered in combination with individual traits and the specific usage context. For instance, study 1's feedback found that users find audio particularly helpful in multitasking. Additionally, study 2's quantitative findings showed that reflection is shaped by both internal preferences and external media exposure. This interplay between personal tendencies and past experiences suggests that reflection is not a one-size-fits-all process. Future research could further explore the tension between these internal and external influences in shaping users’ multimodal preferences.

To optimize reflective nudges, designers and practitioners should consider specific contexts in which modalities are applied. Our findings underscore the need for systematically investigating how individual characteristics impact deliberativeness. While most research focuses on improving deliberation outcomes, less attention has been given to understanding the factors shaping the reflective process itself. By addressing these factors, future studies can develop more targeted strategies to enhance both reflection and deliberativeness.

\subsection{Potential of Utilizing LLMs to Support Self-Reflection}
Unlike traditional online deliberation platforms, where users form opinions through user-driven self-reflection and the consumption of others' comments, our study demonstrates how LLMs can enhance this process using subtle interface nudges. This positions LLMs not just as writing aids but as facilitators of deeper, reflective thinking, steering users toward more informed, thoughtful decisions. Nevertheless, it's crucial to acknowledge that the effectiveness of LLMs relies on the quality of the responses they generate. While we utilized GPT-4.0 along with text-to-image (Microsoft Bing) and text-to-video generation tools (Invideo AI), future advancements in LLM technology, such as HeyGen AI Video Generator or Luma Dream Machine, could yield varied results depending on their specific design and functionality. These innovations will likely further shape how LLMs contribute to deliberation. 

Moreover, the role of LLM-generated content in fostering reflection, compared to human-created nudges, presents an intriguing area for further exploration. Future studies can explore how users engage with and perceive AI-generated versus human-created reflective nudges. 

Lastly, as with the use of any AI or LLM technologies in any domain, inherent risks such as amplifying societal biases must be carefully managed. Generative technologies must be explicitly designed with bias mitigation strategies to prevent perpetuating stereotypes and biases. This requires ongoing transparency about AI's role, clear communication of its limitations, and proactive steps to address potential ethical considerations. 

\subsection{Expanding Modalities for Enhanced Reflection}
Beyond the current straightforward use of modalities, there is potential to extend their application to other stages of the deliberative process.

\paragraph{Checking content before posting.} Incorporating an audio playback of written content by converting textual content into audio would allow users to listen back and validate their understanding before posting. This auditory review could help users ensure that their message aligns with their intended meaning, offering an additional layer of reflection by \textit{hearing} their own thoughts played aloud. 

\paragraph{Usage on the go.} Similarly, providing a text-to-speech feature for other users' opinions would support ``reflection-on-the-go'', particularly for auditory learners. This addition would cater to participants who reflect better by listening or those who prefer a conversational approach over reading, expanding the inclusivity and accessibility of the deliberative process. However, challenges might arise in ensuring that the text-to-speech translation is accurate, particularly in complex or nuanced opinions, which may affect users' comprehension.% For example, auditory reflective users could benefit from hearing nuanced inflections and tones that might not be apparent in text, enhancing their engagement with the content.

\paragraph{Granularity of the modality.} When designing visual aids, the level of granularity --- whether it is single images, image sequences, or multi-message visuals --- plays a key role for visual learners. Tailoring the structure of visual content to user preferences can significantly influence how effective they engage with and reflect on the content. However, a key challenge is balancing the granularity of visual aids such that they are informative without overwhelming the user. Future studies may want to look into the manipulation of modality considerations such as adjusting the granularity of different modalities to create more adaptive reflective experience that is tailored to diverse reflection styles and contexts.

\subsection{Applicability to other Contexts}
While the task presented in this study was specific to an online deliberation context, the multiplicity of opportunities for fostering reflection in other areas, combined with the social affordances provided by multimedia representations, suggests that these approaches hold potential for broader applications. 

\paragraph{Moderation.} In online discussion platforms (e.g., Reddit and Quora), moderators could fine-tune or adjust reflective nudges based on the tone or quality of the discussion, promoting deeper and more respectful engagement. By combining multimodal nudges with moderation, the system becomes more dynamic and adaptable, where nudges are personalized and contextually relevant. This approach also helps prevent excessive dominance by certain perspectives and fosters diverse viewpoints.

\paragraph{Education.} Multimodal reflective nudges can be adapted to educational settings to encourage deeper student engagement and critical thinking. For instance, visualizations or video explanations of complex topics can aid in simplifying abstract concepts. By offering different ways to engage with learning content, students can reflect on their knowledge from multiple angles, promoting active learning and enhanced retention. 

\paragraph{Counseling.} In counseling, using multimodal approaches may help individuals process their emotions and thoughts more effectively. Audio playback of therapeutic dialogues or reflections can help individuals revisit and internalize key insights, while video-based role-playing scenarios can offer practical, real-world examples on how to manage specific situations. These would allow for more personalized and engaging reflections, leading to deeper self-awareness and therapeutic growth.

Future research endeavors could explore the cross-domain generalizability of these findings, seeking to understand the modalities' boundaries and flexibility in diverse domains.
\section{Limitations}
There are limitations in this work. Firstly, study 1 was conducted on young adults. Although this is a common target population for online deliberation platforms, these platforms are widely used in a range of populations. We mitigated this issue in study 2 with a larger and more diverse participants pool. Future work may focus on understanding the cultural aspects of self-reflection when using different modalities in reflective nudges from different populations.

The task employed in both studies was specific, aiming to achieve depth and clarity in understanding the influences of different modalities within a well-defined context~\cite{slack2001establishing}. While this establishes \textbf{internal validity}, it might conflict with the broader goal of generalizability and ecological validity. It is important to note that assessing ecological validity relies on first establishing internal validity~\cite{cahit2015internal, slack2001establishing, campbell2015experimental, cook1979quasi}. Therefore, having a focused investigation on a specific issue allows us to have a detailed examination of the modalities on deliberativeness. Future work could extend the application of the modalities to explore other contentious topics with varying complexity and nature.

Moreover, we utilized the VARK model to identify whether participants have a higher preference of one modality over others. While VARK has its limitations (see section~\ref{sec: learning and reflection}), we solely used the questionnaire for one specific goal: to examine modality preferences rather than making definitive conclusions about learning styles.

Additionally, deliberative processes can vary significantly in duration and complexity. Future studies could benefit from examining longer, multi-phased deliberations that unfold over days or weeks, allowing for deeper exploration of deliberative dynamics.

Lastly, this study investigates two types of reflective nudges — direct and indirect — implemented through persona and storytelling approaches. Our selection was guided by prior research demonstrating the effectiveness of these nudges in comparative evaluations~\cite{yeo2024help}. While other types of reflective nudges may exist, our primary objective was to examine whether different nudge types require distinct modalities to maximize their impact. Our findings underscore that specific modalities align better with particular reflective nudges, offering insights into tailoring nudges for improved reflection and engagement.
\section{Conclusion}
In this work, we investigated how modalities when enacted on different reflective nudges impacts the quality of deliberation. Specifically, we examine how four modalities: text, image, video and audio interact with two types of reflective nudges: direct (persona-based) and indirect (storytelling-based) across two studies. In study 1, we identified text as the subjectively preferred modality for direct reflective nudge while video was favoured for indirect reflective nudge. In study 2, we explored how different modalities can significantly shape deliberativeness. Our results expand current work on self-reflection and online discussions, offering insights into how different modalities support the deliberation process, thus, providing valuable guidance on the use of modalities on online deliberation platforms. In the future, we hope to explore more types of nudges and improve the variety and quality of the content generated with more powerful LLMs.
\section{Acknowledgments}

This work was partially funded by the CGH-SUTD grant CGH-SUTD-HTIF-2022-001.

\bibliographystyle{ACM-Reference-Format}
\bibliography{reference.bib}

\newpage
\section*{Appendix}
The appendix contains supplementary data, which, while not part of the analysis, may provide additional details that can help in replicating the study. 

\subsection*{Prompt Engineering}
\label{sec: prompt engineering}
Tables~\ref{tab: prompt engineering} and \ref{tab: prompt constraint rationale} detail the prompts used in GPT and AI generative models, along with the specific constraints applied for each modality. According to the ChatGPT API specifications, the ``system role'' defines the model's assumed role in the session, while ``user input'' provides the prompt guiding the model's response.

\subsection*{Demographics Profile}
We collated participants' topic knowledge and topic interest (TK-TI) scores, their self-reported self-reflection and insight scale (SRIS) scores, VARK scores and external exposures and interactions with the different modalities in both reflective nudge (see Tables~\ref{tab: st1-demo} and \ref{tab: st2-demo}), to account for any potential fixed effects associated with these four covariates. 

In study 1, it's worth noting that all four covariates showed no statistically significant impact on the rankings of the modalities. Furthermore, we found no meaningful correlations between any of the four covariates and their interactions with any of the nudges.

In study 2, we report the covariates that are significant in section~\ref{sec: quantitative}.

\subsection*{Summary of Qualitative Feedback for Study 1}
\begin{figure*}[!htbp]
  \centering
  \includegraphics[width=\linewidth]{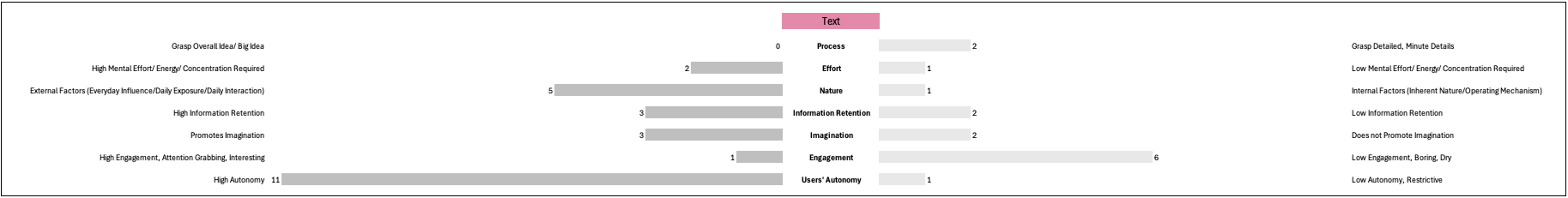}
  \caption{Butterfly Chart for the Text Modality across Both Nudge Types}
  \label{fig: butterfly chart text}
  \Description{Butterfly chart showing how the text modality fare for each of the categories under Kahneman's dual-system thinking model.}
\end{figure*}

\begin{figure*}[!htbp]
  \centering
  \includegraphics[width=\linewidth]{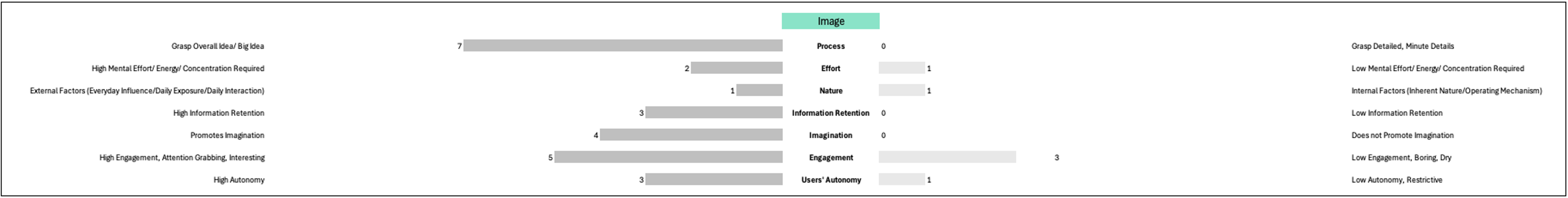}
  \caption{Butterfly Chart for the Image Modality across Both Nudge Types}
  \label{fig: butterfly chart image}
  \Description{Butterfly chart showing how the image modality fare for each of the categories under Kahneman's dual-system thinking model.}
\end{figure*}

\begin{figure*}[!htbp]
  \centering
  \includegraphics[width=\linewidth]{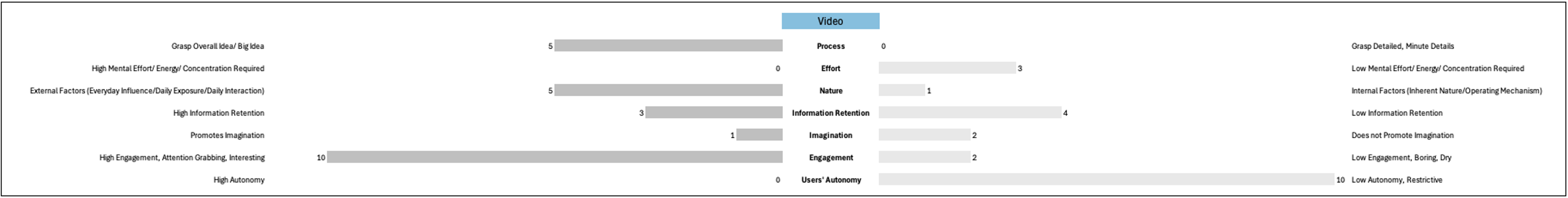}
  \caption{Butterfly Chart for the Video Modality across Both Nudge Types}
  \label{fig: butterfly chart video}
  \Description{Butterfly chart showing how the video modality fare for each of the categories under Kahneman's dual-system thinking model.}
\end{figure*}

\begin{figure*}[!htbp]
  \centering
  \includegraphics[width=\linewidth]{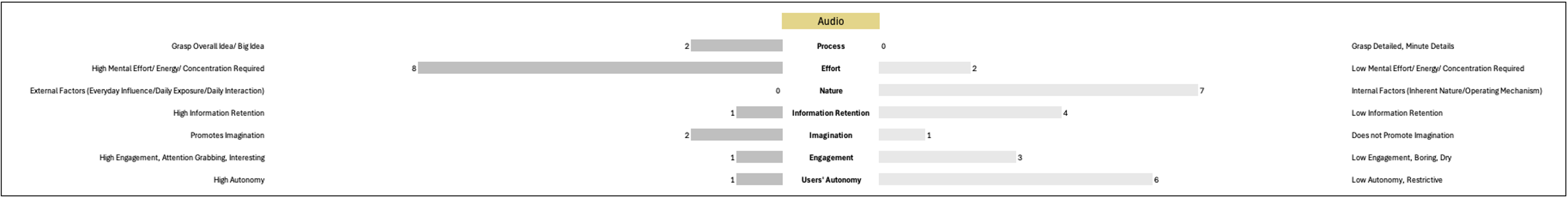}
  \caption{Butterfly Chart for the Audio Modality across Both Nudge Types}
  \label{fig: butterfly chart audio}
  \Description{Butterfly chart showing how the audio modality fare for each of the categories under Kahneman's dual-system thinking model.}
\end{figure*}

Qualitative feedback for study 1 for both reflective nudges across the four modalities are summarized in Figures~\ref{fig: butterfly chart text} to \ref{fig: butterfly chart audio}. Feedback is classified using Kahneman's dual system thinking model~\cite{kahneman2002maps} to illustrate that the same modalities can produce varying or even contradictory results depending on the type of reflective nudge applied.

\newpage

\begin{table*}[!htbp]
\caption{The prompts utilized to generate the modalities along with the corresponding constraints}
\label{tab: prompt engineering}
\scalebox{0.6}{
\begin{tabular}{|l|lll|l|}
\hline
\textbf{Reflective Nudges} &
  \multicolumn{1}{l|}{\textbf{Prompt Template for Text Modality}} &
  \multicolumn{1}{l|}{\textbf{Prompt Template for Image Modality}} &
  \textbf{Prompt Template for Video Modality} &
  \textbf{Prompt Template for Audio Modality} \\ \hline
\textit{\begin{tabular}[c]{@{}l@{}}For both reflective \\ nudges\end{tabular}} &
  \multicolumn{3}{l|}{\begin{tabular}[c]{@{}l@{}}\textbf{system role:} You are a helpful assistant focusing on supporting users' self-reflection on a given topic.\\      \\ \textbf{User input:} Topic: {[}topic{]}.\end{tabular}} &
  \multirow{3}{*}{\begin{tabular}[c]{@{}l@{}}Audio prompts are not required for this \\ process, as the text generated in the text \\ modality is directly input into the text-to-\\ speech AI tool. The tool automatically \\ converts the provided text into audio, \\ generating the narration based on the \\ content supplied. We manually \\ selected a gender-appropriate voice that \\ matches the gender as specified in the \\ text modality.\end{tabular}} \\ \cline{1-4}
\begin{tabular}[c]{@{}l@{}}Direct Reflective \\ Nudge (Persona)\end{tabular} &
  \multicolumn{1}{l|}{\begin{tabular}[c]{@{}l@{}}For the above topic, create ten distinct personas \\ representing different perspectives on the topic. \\ Provide the name, age and occupation for each \\ persona. \\      \\ Here is the format of the results: \\ 1. {[}Name1{]}, {[}Age1{]}, {[}Occupation1{]}, {[}Perspective1{]}\\ 2. {[}Name2{]}, {[}Age2{]}, {[}Occupation2{]}, {[}Perspective2{]}\\      ...\\      \\ Requirements:\\ Create five male personas and five female personas;\\ For each perspective, be concise, giving at most three \\ sentences;\\ No duplicates;\\ Ten distinct versions only\end{tabular}} &
  \multicolumn{1}{l|}{\begin{tabular}[c]{@{}l@{}}\textit{For each persona generated in the text} \\ \textit{modality, we use the following prompt to} \\ \textit{generate the corresponding image via AI:}\\      \\ Persona: {[}persona1{]}     \\ Create a photorealistic image with realistic \\ textures and lighting of the persona. The\\ background should be related to the \\ occupation of the persona. \\      \\ Requirements: \\ Age, gender and occupation should follow \\ the persona. The style is photorealistic and \\ not cartoonish. The image has no wordings.\end{tabular}} &
  \begin{tabular}[c]{@{}l@{}}\textit{For each persona generated in the text} \\ \textit{modality, we input the entire script into} \\ \textit{the AI.} \textit{We then use the following prompt} \\ \textit{to generate the corresponding video:}\\      \\ Persona: {[}persona1{]}     \\ Create a photorealistic video featuring the \\ persona. The background should be realistic \\ and aligned with the persona’s occupation \\ and worldview, with natural lighting and no \\ cartoonish elements. The voice narration \\ should match the persona’s gender and follow \\ the script exactly as provided, with no \\ additional text. The video should be visually \\ compelling and immersive, showcasing \\ a background relevant to the persona’s \\ occupation or environment. \\      \\ Requirements: \\ Resolution: 1080p\\ Audience: Relevant to the persona’s occupation\\ Style: Photorealistic, professional\\ Platform: YouTube Shorts or Instagram Reels\\ Script: Follow the provided persona text exactly \\ without any modifications.\end{tabular} &
   \\ \cline{1-4}
\begin{tabular}[c]{@{}l@{}}Indirect Reflective \\ Nudge (Storytelling)\end{tabular} &
  \multicolumn{1}{l|}{\begin{tabular}[c]{@{}l@{}}Following the ten personas created earlier, generate a \\ story for each of the persona on the topic matter. \\      \\ Here is the format of the results:\\ Story1:\\ Story2:\\ …\\      \\ Requirements:\\ Create five stories with a positive tone and \\ five stories with a negative tone;\\ No duplicates or similar story line;\\ Ten distinct versions only\\      \\ * Depending on the length of the story generated, \\ we prompt GPT to either lengthen (Extend Story1) \\ or condense (Make Story1 concise) the narrative.\end{tabular}} &
  \multicolumn{1}{l|}{\begin{tabular}[c]{@{}l@{}}\textit{For each story generated in the text modality,} \\ \textit{we prompt AI to create images that visually} \\ \textit{represent key moments in the narrative. We} \\ \textit{generate one image at a time, which are then} \\ \textit{combined to form a cohesive visual }\\ \textit{representation of the entire story.}\\      \\ Section of the story: {[}story section1{]}.     \\ Create a photorealistic image with realistic \\ textures and lighting of the section of the story. \\ The background should be related to the \\ occupation of the character in the story. \\ \\      \\ Requirements: \\ Age, gender and occupation should follow \\ the character in the story.\\ The style is photorealistic and not cartoonish. \\ The image has no wordings. \\ \textasciicircum The style should be consistent with the \\ previous image.\\      \\ \textasciicircum Only for subsequent images generated \\ by the AI to maintain visual coherence.\end{tabular}} &
  \begin{tabular}[c]{@{}l@{}}\textit{For each story generated in the text modality, }\\ \textit{we input the entire script into the AI.} \\ \textit{Pauses were manually added to ensure natural }\\ \textit{pacing and alignment with the story's speech }\\ \textit{patterns. We then use the following prompt to }\\ \textit{generate the corresponding video:}\\      \\ Story: {[}story1{]}     \\ Create a photorealistic video featuring the \\ story. The background should be realistic \\ and aligned with the character's occupation and \\ worldview, with natural lighting and no cartoonish \\ elements. The voice narration should match the \\ character's gender and follow the script exactly as \\ provided, with no additional text. The video should \\ be visually compelling and immersive, showcasing \\ a background relevant to the character's \\ occupation or environment. \\      \\ Requirements: \\ Resolution: 1080p\\ Audience: Relevant to the character's occupation\\ Style: Photorealistic, professional\\ Platform: YouTube Shorts or Instagram Reels\\ Script: Follow the provided text exactly without any \\ modifications.\end{tabular} &
   \\ \hline
\end{tabular}}
\Description{This table details the prompts used to generate the variants for each of the modalities, along with the corresponding constraints applied when communicating with the ChatGPT API and AI generative models. These prompts adhere to the template established by White et al., which entails defining a task, incorporating constraints, and setting clear expectations for the generated output.}
\end{table*}

\newpage

\begin{table*}[!htbp]
\caption{Rationales of the constraints set out for GPT and the AI Generative tools}
\label{tab: prompt constraint rationale}
\centering
\begin{tblr}{
  width = \textwidth,
  colspec = {Q[352]Q[588]},
  row{1} = {c},
  hlines,}
\textbf{Constraints} & \textbf{Rationale} \\
{(Direct Reflective Nudge: Persona) Create five male personas and five female personas. \\~} & {\labelitemi\hspace{\dimexpr\labelsep+0.5\tabcolsep}Scholars have found that large language models such as GPT-3 produce gender stereotypes and biases~\cite{brown2020language,lucy2021gender, huang2019reducing, nozza2021honest, johnson2022ghost} \\\labelitemi\hspace{\dimexpr\labelsep+0.5\tabcolsep} Hence, this constraint was set out to mitigate any potential gender imbalances as the discussion topic is a contentious one. Moreover, GPT was specifically instructed to assign gender-neutral names to the personas. This approach prevents the association of particular names with male or female identities, fostering a more equitable and unbiased representation. \\\labelitemi\hspace{\dimexpr\labelsep+0.5\tabcolsep} Ensures GPT does not provide viewpoints that gives preferential treatment of one gender over the other.} \\
(Direct Reflective Nudge: Persona) For each perspective, be concise, giving at most three sentences. & {\labelitemi\hspace{\dimexpr\labelsep+0.5\tabcolsep}This constraint was introduced after observing GPT's tendency to generate long perspectives during testing. \\\labelitemi\hspace{\dimexpr\labelsep+0.5\tabcolsep} Ensures GPT deliver concise perspectives.} \\
(Indirect Reflective Nudge: Storytelling) Of the ten stories, create five with a positive tone and five with a negative tone. & \labelitemi\hspace{\dimexpr\labelsep+0.5\tabcolsep} Ensures that the reflector captures a broad spectrum of impacts, including both positive and negative aspects to avert one-sidedness. \\
(Indirect Reflective Nudge: Storytelling) Extend or make concise. & {\labelitemi\hspace{\dimexpr\labelsep+0.5\tabcolsep} Ensures a well-rounded representation of stories --- three long stories, three short stories, and four stories of medium length. \\\labelitemi\hspace{\dimexpr\labelsep+0.5\tabcolsep} This strategy allows us to capture the full spectrum of narrative possibilities and thereby provides a more thorough analysis of the reflector's effectiveness.} \\
(Image Modality) Image has no wordings. & {\labelitemi\hspace{\dimexpr\labelsep+0.5\tabcolsep} AI-generated images often produce words that are not human-readable, as the algorithms focus on replicating the visual shapes of letters and numbers rather than rendering actual text. This issue is similar to the challenges AI faces in accurately generating human hands, which frequently results in distorted representations due to the complexity of shapes involved~\cite{keyes2023hands}.} \\
(Video Modality) Platform: YouTube Shorts or Instagram Reels. & {\labelitemi\hspace{\dimexpr\labelsep+0.5\tabcolsep} Ensures that the generated video aligns with the format requirements of popular media platforms such as TikTok, Instagram Reels, and YouTube Shorts. By adhering to these guidelines, the video remains concise, typically between 90-120 seconds, in line with the standard duration of videos commonly found on these platforms.} \\
(Video Modality) Script: Follow the provided text exactly without any modifications. & {\labelitemi\hspace{\dimexpr\labelsep+0.5\tabcolsep} During testing, AI video generation platforms often augment scripts by adding extra narrative elements to enhance immersion and provide more context. To maintain consistency across all modalities, we explicitly constrain the AI to avoid any content alterations, ensuring that the script remains unchanged throughout the different formats.} \\
Here is the format of the results & \labelitemi\hspace{\dimexpr\labelsep+0.5\tabcolsep} Ensures that GPT provides results in a consistent format.      
\end{tblr}
\Description{This table provides a breakdown of the constraints applied to GPT and AI generative models during the generation of the variants for each modality, along with the underlying rationales for each constraint. The table has two columns with column header: constraints and rationale. In the constraints column, we list the specific constraints implemented for each modality and reflective nudge. Meanwhile, the rationale column elucidates the motivations behind the establishment of these constraints.}
\end{table*}

\newpage

\begin{table*}[!htbp]
\caption{Demographic Profiles of Participants in Study 1}
\label{tab: st1-demo}
\scalebox{0.74}{
\begin{tabular}{|ll|c|c|}
\hline
\multicolumn{1}{|l|}{\textbf{Category}} &
  \textbf{Dimensions} &
  \textbf{Direct Reflective Nudge (Persona)} &
  \textbf{Indirect Reflective Nudge (Storytelling)} \\ \hline
\multicolumn{2}{|l|}{Total   Number of Participants} &
  10 &
  10 \\ \hline
\multicolumn{1}{|l|}{\multirow{2}{*}{Gender}} &
  Total Number of Males &
  4 &
  4 \\ \cline{2-4} 
\multicolumn{1}{|l|}{} &
  Total Number of Females &
  6 &
  6 \\ \hline
\multicolumn{1}{|l|}{Age} &
  Average Age &
  23.9 &
  24.1 \\ \hline
\multicolumn{1}{|l|}{\multirow{2}{*}{Ethnicity}} &
  Asian or Pacific Islander &
  10 &
  9 \\ \cline{2-4} 
\multicolumn{1}{|l|}{} &
  Hispanic or Latino &
  0 &
  1 \\ \hline
\multicolumn{1}{|l|}{\multirow{2}{*}{Education}} &
  Bachelor's Degree &
  8 &
  8 \\ \cline{2-4} 
\multicolumn{1}{|l|}{} &
  Post-Graudate Degree &
  2 &
  2 \\ \hline
\multicolumn{1}{|l|}{\multirow{2}{*}{Language}} &
  English is first language &
  7 &
  8 \\ \cline{2-4} 
\multicolumn{1}{|l|}{} &
  English is NOT first language &
  3 &
  2 \\ \hline
\multicolumn{2}{|l|}{TITK Score ($M \pm S.D.$)} &
  $20.10 \pm 5.07$ &
  $18.70 \pm 3.92$ \\ \hline
\multicolumn{2}{|l|}{SRIS Score ($M \pm S.D.$)} &
  $89.00 \pm 11.07$ &
  $86.80 \pm 10.05$ \\ \hline
\multicolumn{1}{|l|}{\multirow{4}{*}{VARK Model (Internal -   Inherent Reflecting Styles)}} &
  Visual (V) Score ($M \pm S.D.$) &
  $8.70 \pm 2.71$ &
  $9.10 \pm 3.31$ \\ \cline{2-4} 
\multicolumn{1}{|l|}{} &
  Audio (A) Score ($M \pm S.D.$) &
  $7.70 \pm 3.30$ &
  $5.10 \pm 2.18$ \\ \cline{2-4} 
\multicolumn{1}{|l|}{} &
  Read/Write (R) Score ($M \pm S.D.$) &
  $6.80 \pm 3.71$ &
  $7.00 \pm 3.46$ \\ \cline{2-4} 
\multicolumn{1}{|l|}{} &
  Kinesthetic (K) Score ($M \pm S.D.$) &
  $10.90 \pm 2.64$ &
  $9.70 \pm 2.26$ \\ \hline
\multicolumn{1}{|l|}{\multirow{4}{*}{Daily Usage (External -  Influences and Interactions)}} &
  \begin{tabular}[c]{@{}l@{}}Frequency of Exposure to Text:\\ Blog Posts, Online Articles, News \\ ($M \pm S.D.$)\end{tabular} &
  $4.20 \pm 0.79$ &
  $3.30 \pm 1.25$ \\ \cline{2-4} 
\multicolumn{1}{|l|}{} &
  \begin{tabular}[c]{@{}l@{}}Frequency of Exposure to Images: \\ Instagram Posts, Facebook Posts, Pinterest\\ ($M \pm S.D.$)\end{tabular} &
  $4.00 \pm 0.94$ &
  $3.40 \pm 1.43$ \\ \cline{2-4} 
\multicolumn{1}{|l|}{} &
  \begin{tabular}[c]{@{}l@{}}Frequency of Exposure to Video:\\ Instagram Reels, Youtube Shorts, TikTok\\ ($M \pm S.D.$)\end{tabular} &
  $4.60 \pm 0.97$ &
  $4.10 \pm 1.45$ \\ \cline{2-4} 
\multicolumn{1}{|l|}{} &
  \begin{tabular}[c]{@{}l@{}}Frequency of Exposure to Audio: \\ Podcasts ($M \pm S.D.$)\end{tabular} &
  $2.20 \pm 1.23$ &
  $2.50 \pm 1.18$ \\ \hline
\multicolumn{2}{|l|}{Average Completion Time in Minutes} &
  45.6 &
  48.0 \\ \hline
\end{tabular}}
\Description{This table provides a comprehensive overview of the demographic characteristics of the participants in Study 1. Demographic profile captures Participants, Gender, Age, Education, TK-TI Scores, SRIS Scores, VARK scores and Daily Interactions with the modalities.}
\end{table*}

\begin{table*}[!htbp]
\caption{Demographic Profiles of Participants in Study 2}
\label{tab: st2-demo}
\scalebox{0.5}{
\begin{tabular}{|ll|c|c|c|c|c|c|c|c|}
\hline
\multicolumn{1}{|l|}{\textbf{Category}} &
  \textbf{Dimensions} &
  \textbf{Persona (Text)} &
  \textbf{Persona (Image)} &
  \textbf{Persona (Video)} &
  \textbf{Persona (Audio)} &
  \textbf{Storytelling (Text)} &
  \textbf{Storytelling (Image)} &
  \textbf{Storytelling (Video)} &
  \textbf{Storytelling (Audio)} \\ \hline
\multicolumn{2}{|l|}{Total Number of Participants} &
  25 &
  25 &
  25 &
  25 &
  25 &
  25 &
  25 &
  25 \\ \hline
\multicolumn{1}{|l|}{\multirow{3}{*}{Gender}} &
  Total Number of Males &
  14 &
  11 &
  11 &
  12 &
  13 &
  11 &
  15 &
  18 \\ \cline{2-10} 
\multicolumn{1}{|l|}{} &
  Total Number of Females &
  10 &
  14 &
  14 &
  13 &
  12 &
  14 &
  10 &
  7 \\ \cline{2-10} 
\multicolumn{1}{|l|}{} &
  Prefer Not to Say &
  1 &
  0 &
  0 &
  0 &
  0 &
  0 &
  0 &
  0 \\ \hline
\multicolumn{1}{|l|}{Age} &
  Average Age &
  30.72 &
  34.28 &
  32.48 &
  35.84 &
  33.68 &
  33.44 &
  37.32 &
  36.52 \\ \hline
\multicolumn{1}{|l|}{\multirow{7}{*}{Ethnicity}} &
  Asian or Pacific Islander &
  3 &
  2 &
  0 &
  2 &
  0 &
  0 &
  3 &
  3 \\ \cline{2-10} 
\multicolumn{1}{|l|}{} &
  Black or African American &
  0 &
  0 &
  1 &
  1 &
  1 &
  0 &
  0 &
  0 \\ \cline{2-10} 
\multicolumn{1}{|l|}{} &
  Hispanic or Latino &
  0 &
  1 &
  1 &
  0 &
  0 &
  0 &
  0 &
  1 \\ \cline{2-10} 
\multicolumn{1}{|l|}{} &
  Multiracial or Biracial &
  0 &
  0 &
  0 &
  0 &
  0 &
  0 &
  0 &
  0 \\ \cline{2-10} 
\multicolumn{1}{|l|}{} &
  Native American or Alaska Native &
  3 &
  0 &
  1 &
  1 &
  0 &
  1 &
  0 &
  0 \\ \cline{2-10} 
\multicolumn{1}{|l|}{} &
  White or Caucasian &
  18 &
  22 &
  22 &
  21 &
  24 &
  24 &
  22 &
  21 \\ \cline{2-10} 
\multicolumn{1}{|l|}{} &
  Other &
  1 &
  0 &
  0 &
  0 &
  0 &
  0 &
  0 &
  0 \\ \hline
\multicolumn{1}{|l|}{\multirow{5}{*}{Education}} &
  No Diploma or less &
  0 &
  0 &
  0 &
  0 &
  0 &
  0 &
  0 &
  0 \\ \cline{2-10} 
\multicolumn{1}{|l|}{} &
  High School, Diploma or the equivalent &
  3 &
  1 &
  2 &
  2 &
  0 &
  1 &
  0 &
  2 \\ \cline{2-10} 
\multicolumn{1}{|l|}{} &
  Associate's Degree &
  1 &
  1 &
  1 &
  2 &
  1 &
  1 &
  0 &
  3 \\ \cline{2-10} 
\multicolumn{1}{|l|}{} &
  Bachelor's Degree &
  15 &
  23 &
  19 &
  20 &
  23 &
  18 &
  20 &
  18 \\ \cline{2-10} 
\multicolumn{1}{|l|}{} &
  Post-graduate Degree &
  6 &
  0 &
  3 &
  1 &
  1 &
  5 &
  5 &
  2 \\ \hline
\multicolumn{1}{|l|}{\multirow{2}{*}{Language}} &
  English is first language &
  23 &
  25 &
  24 &
  25 &
  25 &
  25 &
  24 &
  25 \\ \cline{2-10} 
\multicolumn{1}{|l|}{} &
  English is NOT first language &
  2 &
  0 &
  1 &
  0 &
  0 &
  0 &
  1 &
  0 \\ \hline
\multicolumn{2}{|l|}{TITK Score ($M \pm S.D.$)} &
  $21.04 \pm 4.89$ &
  $23.16 \pm 3.50$ &
  $22.28 \pm 5.26$ &
  $22.28 \pm 6.78$ &
  $23.20 \pm 4.37$ &
  $23.20 \pm 5.98$ &
  $27.08 \pm 5.99$ &
  $26.12 \pm 5.90$ \\ \hline
\multicolumn{2}{|l|}{SRIS Score ($M \pm S.D.$)} &
  $81.32 \pm 11.91$ &
  $84.84 \pm 9.48$ &
  $84.64 \pm 13.73$ &
  $85.64 \pm 12.68$ &
  $84.68 \pm 18.45$ &
  $86.24 \pm 13.82$ &
  $83.20 \pm 12.43$ &
  $82.76 \pm 10.33$ \\ \hline
\multicolumn{1}{|l|}{\multirow{4}{*}{VARK Model (Internal -   Inherent Reflecting Styles)}} &
  Visual (V) Score ($M \pm S.D.$) &
  $7.52 \pm 3.11$ &
  $8.40 \pm 3.72$ &
  $7.92 \pm 3.98$ &
  $6.92 \pm 3.59$ &
  $7.68 \pm 3.87$ &
  $5.40 \pm 2.47$ &
  $6.12 \pm 2.71$ &
  $5.88 \pm 3.23$ \\ \cline{2-10} 
\multicolumn{1}{|l|}{} &
  Audio (A) Score ($M \pm S.D.$) &
  $8.96 \pm 3.85$ &
  $7.36 \pm 2.84$ &
  $8.48 \pm 3.63$ &
  $7.36 \pm 4.19$ &
  $8.28 \pm 3.68$ &
  $7.88 \pm 3.10$ &
  $8.64 \pm 2.96$ &
  $6.80 \pm 2.33$ \\ \cline{2-10} 
\multicolumn{1}{|l|}{} &
  Read/Write (R) Score ($M \pm S.D.$) &
  $7.96 \pm 3.51$ &
  $7.68 \pm 3.20$ &
  $8.16 \pm 2.78$ &
  $7.36 \pm 3.77$ &
  $7.48 \pm 3.79$ &
  $6.92 \pm 2.41$ &
  $6.32 \pm 2.76$ &
  $6.12 \pm 3.49$ \\ \cline{2-10} 
\multicolumn{1}{|l|}{} &
  Kinesthetic (K) Score ($M \pm S.D.$) &
  $8.44 \pm 3.50$ &
  $9.00 \pm 3.35$ &
  $8.08 \pm 2.71$ &
  $7.72 \pm 3.66$ &
  $8.52 \pm 3.20$ &
  $6.84 \pm 3.10$ &
  $8.32 \pm 2.41$ &
  $7.20 \pm 2.72$ \\ \hline
\multicolumn{1}{|l|}{\multirow{4}{*}{Daily Usage (External -   Influences and Interactions)}} &
  \begin{tabular}[c]{@{}l@{}}Frequency of Exposure to Text: \\ Blog Posts, Online Articles, News\\ ($M \pm S.D.$)\end{tabular} &
  $3.72 \pm 1.10$ &
  $3.80 \pm 0.65$ &
  $3.84 \pm 0.99$ &
  $3.92 \pm 0.91$ &
  $3.76 \pm 0.72$ &
  $3.72 \pm 0.98$ &
  $4.20 \pm 0.71$ &
  $4.24 \pm 0.88$ \\ \cline{2-10} 
\multicolumn{1}{|l|}{} &
  \begin{tabular}[c]{@{}l@{}}Frequency of Exposure to Images: \\ Instagram Posts, Facebook   Posts, Pinterest\\ ($M \pm S.D.$)\end{tabular} &
  $4.28 \pm 0.74$ &
  $4.24 \pm 0.93$ &
  $4.24 \pm 0.88$ &
  $4.36 \pm 1.11$ &
  $4.04 \pm 0.93$ &
  $4.04 \pm 1.06$ &
  $4.68 \pm 0.48$ &
  $4.32 \pm 0.99$ \\ \cline{2-10} 
\multicolumn{1}{|l|}{} &
  \begin{tabular}[c]{@{}l@{}}Frequency of Exposure to Video:\\ Instagram Reels, Youtube Shorts, TikTok\\ ($M \pm S.D.$)\end{tabular} &
  $4.20 \pm 0.91$ &
  $4.56 \pm 0.87$ &
  $4.08 \pm 0.95$ &
  $4.44 \pm 0.96$ &
  $4.40 \pm 0.76$ &
  $4.16 \pm 0.99$ &
  $4.64 \pm 0.70$ &
  $4.44 \pm 0.92$ \\ \cline{2-10} 
\multicolumn{1}{|l|}{} &
  \begin{tabular}[c]{@{}l@{}}Frequency of Exposure to Audio: \\ Podcasts ($M \pm S.D.$)\end{tabular} &
  $3.76 \pm 1.16$ &
  $3.52 \pm 1.08$ &
  $3.60 \pm 1.15$ &
  $3.68 \pm 1.07$ &
  $3.56 \pm 1.00$ &
  $3.56 \pm 1.04$ &
  $3.92 \pm 1.00$ &
  $3.88 \pm 1.10$ \\ \hline
\multicolumn{2}{|l|}{Average Completion Time in Minutes} &
  25.66 &
  47.22 &
  25.50 &
  32.18 &
  28.31 &
  32.88 &
  24.34 &
  24.96 \\ \hline
\end{tabular}}
\Description{This table provides a comprehensive overview of the demographic characteristics of the participants in Study 2. Demographic profile captures Participants, Gender, Age, Education, TK-TI Scores, SRIS Scores, VARK scores and Daily Interactions with the modalities.}
\end{table*}

\clearpage
\newpage

\end{document}